\documentclass{aa}

\usepackage{newtxtext,newtxmath}
\usepackage[T1]{fontenc}

\usepackage{pdflscape}

\usepackage[dvipsnames]{xcolor}
\usepackage[normalem]{ulem}
\usepackage[bottom]{footmisc}
\usepackage{lineno}

\usepackage{savesym}
\usepackage[nointegrals]{wasysym}
\usepackage{color}
\usepackage[normalem]{ulem}

\usepackage[hang]{subfigure}
\usepackage{hyperref}
\usepackage{accents}
\usepackage{threeparttable, booktabs}

\savesymbol{tablenum}
\usepackage{physics}
\usepackage{siunitx}
\usepackage{mathtools}
\restoresymbol{SIX}{tablenum}

\usepackage{changes}

\DeclareSIUnit{\erg}{erg}
\DeclareSIUnit{\jansky}{Jy}
\DeclareSIUnit{\parsec}{pc}
\DeclareSIUnit{\yr}{yr}
\DeclareSIUnit\msun{M\textsubscript{\astrosun}}
\DeclareSIUnit{\radian}{rad}
\DeclareSIUnit{\pixel}{px}
\DeclareSIUnit{\day}{d}
\DeclareSIUnit{\arcsectxt}{as}
\DeclareSIUnit{\arcsec}{arcsec}
\DeclareSIUnit{\correlation}{correlation}
\DeclareSIUnit{\strain}{strain}

\newcommand{\mtx}[1]{\mathbf{#1}}

\newcommand{\gw}{\mathrm{gw}}
\newcommand{\GW}{\mathrm{GW}}
\newcommand{\GWB}{\mathrm{GWB}}

\newcommand{\mc}{\mathcal{M}_\mathrm{c}}
\newcommand{\lmc}{\log_{10}\mathcal{M}_\mathrm{c}}

\begin{document} 

\title{The role of distant pulsars in the detectability of continuous gravitational waves}

\author{Kathrin Grunthal \inst{1}\thanks{\email{kgrunthal@mpifr-bonn.mpg.de}}
          \and
          Nataliya Porayko \inst{2,1}
          \and
          David J. Champion \inst{1}
          \and
          Michael Kramer \inst{1,3}
          }

\institute{$^1$Max-Planck-Institut für Radioastronomie, Auf dem Hügel 69, D-53121 Bonn, Germany\\
$^2$Sternberg Astronomical Institute, Moscow State University, Universitetsky pr., 13, Moscow 119234, Russia\\
$^3$Jodrell Bank Centre for Astrophysics, University of Manchester,
 Department of Physics and Astronomy, Alan-Turing Building, Oxford Street, Manchester M13 9PL, UK
}

   \date{\today}


\abstract
   {One of the imminent science goals of pulsar timing arrays (PTAs) is the detection of a continuous gravitational wave (CGW) emitted by an individual supermassive black hole binary (SMBHB). SMBHBs that cause CGWs with GW frequencies $f_\GW > \SI{10}{\nano\hertz}$ (high-frequency end of the nano-Hertz GW spectrum) have undergone  significant orbital evolution, hence a change of $f_\GW$ over time. In PTA data sets with sufficiently long observational time span, this means that the Earth and pulsar terms' contributions to the CGW signal signature can eventually become resolvable. Since the pulsar term is accumulated incoherently and thus often treated as an additional source of noise, this separation can prove to be beneficial for the detection of the CGW signal in the PTA data set. }
   {We aim to investigate to what extent resolvable Earth and pulsar terms affect currently used techniques for CGW searches with PTA data sets, that treat the pulsar term as an additional source noise. We focus on the dependency of the pulsar term frequencies on the pulsar's distance. We aim to answer the question of whether adding more distant pulsars to a PTA data set can mitigate biases and improve the detection of CGWs.
   }
   {We use simulated PTA data sets based on the EPTA DR2 and IPTA DR2 pulsars in order to study the performance of the Earth-term-only Bayesian parameter estimation of the circular SMBHB model parameters and the frequentist narrow-band optimal statistic in the light of resolved pulsar terms due to larger pulsar distances.}
    {We show that under ideal conditions, more distant pulsars can facilitate the CGW search with PTA data sets. Bayesian parameter estimation is yielding better parameter constraints and the frequentist search becomes more stable. However, using the realistic data set simulations, it was found that other configuration parameters of a PTA, such as the anisotropic distribution of pulsars and the effective number of pulsars in a PTA, can play a crucial role to the importance of this effect.}
   {}

   \maketitle

\section{Introduction}

The direct detection of a gravitational wave (GW) originating from the interaction of a black-hole binary allows us to draw conclusions about the masses of components, its orientation, distance and eccentricity by analysing the measured wave form \citep{Maggiore_2000}. Results of GW searches in the past decade were dominated by the detection of GWs in the high-frequency regime (\SIrange{1}{1000}{\hertz}) caused by merging stellar mass black-hole binary using ground-based interferometers such as LIGO, VIRGO and KAGRA \citep{Bailes_Rev2021}. Complementing the high-frequency GW regime, significant efforts were made over the past decades to also detect GWs at lower frequencies ($f_\GW \sim \si{\nano\hertz}$). These low-frequency GWs can originate from supermassive black hole (SMBH) binaries \citep{Rajagopal_1995}. These SMBHs are expected to be found in the centres of galaxies \citep{Begelman_1980,Kormendy_1995,Milosavljevic_2001}, thus their GW fingerprint can serve as a tracer of galaxy interaction and formation history \citep{BurkeSpolaor_2015}. To detect these low-frequency GWs, a galaxy-sized GW detector is needed. 

The so-called Pulsar Timing Arrays (PTAs) are providing this kind of detector by exploiting the highly predictable rotational behaviour of millisecond pulsars \citep{EstabrookWahlquist_1978,Sazhin_1978,Detweiler_1979,Foster_1990}. Pulsar timing describes the method of exploiting the extraordinary rotational stability of millisecond pulsars by precisely recording their pulse arrival times (ToAs) at the radio telescope using maser clocks tied to international time standards, and comparing the measured time series to the one predicted by the timing model, e.g.\ \cite{handbook}. The deviations between measurement and model, the so-called timing residuals are then used to determine the model parameters, such as astrometric or orbital parameters, to unmatched precision. With typical observation time spans of the order of tens of years, PTA data sets allow for the analysis of periodic signals ranging in frequencies from $f_\mathrm{min} = 1/T_\mathrm{obs} \sim \si{\nano\hertz}$ to the Nyquist frequency $1/(2\Delta t_\mathrm{obs})$ determined by the observational cadence of the PTA. A typical cadence of monthly observations sets the upper signal frequency limit of a PTA to $\sim \SI{100}{\nano\hertz}$. 

By combining the ToA measurements and timing models of several pulsars, it is possible to search for the pulsar residuals induced by GWs in the nano-Hertz regime. The degree of correlation of the GW-induced delays, the overlap reduction function, follows, the Hellings-Downs' (HD) curve \citep{HellingsDowns_1983} as a function of the pulsars' angular separation, under the assumption of GR. This correlation function is the fingerprint of the presence of GWs in the data set. The GW analysis of a PTA is built upon the Fourier decomposition of the data set, where the Fourier frequencies are multiples of $f_\mathrm{min}$, also referred to as ``frequency bins''.

The two major GW signals that are searched for using PTAs are (i) the stochastic gravitational wave background (GWB), which manifests as a stochastic common red noise signal in the pulsar timing residuals \citep[see e.g.][]{AllenRomano_1999,Phinney_2001,Jaffe_2003}, and (ii) continuous gravitational waves (CGWs) originating from individual SMBHB inspirals, which produce a deterministic correlated wave-like pattern in the residuals \citep{Foster_1990,SesanaVecchio_2010,Ravi_2012}. Both signals are expected to be HD-correlated \citep{RomanoAllen_2024}.

Based on realistic simulations of cosmological population of SMBHs, it was found that the first source that is expected to be detected in the PTA band is the GWB \citep{Rosado_2015}. In June 2023, various PTA collaborations around the world published compelling evidence for the presence of an HD-correlated common red noise signal in their respective data sets \citep{EPTA_DR2_GWB,PPTA_GWB,NG15_GWB,CPTA_GWB}. With this, PTAs are undoubtedly entering into the detection era, in which the imminent detection of the GWB puts a renewed focus on the search for a CGW signal. Additionally, the ongoing combination of the different regional PTA data sets under the umbrella of the international Pulsar Timing Array (IPTA) \citep{IPTA_2013}, as well as the contribution of modern radio telescopes such as the MeerKAT radio telescope \citep{Spiewak_2022} steadily increases the sensitivity and resolution of PTAs \citep{Babak_2024}. Building upon earlier works \citep[e.g.][]{Sesana_2004,SesanaVecchioVolonteri_2009,SesanaVecchio_2010,BabakSesana2012,Petiteau_2013}, this raises again the question of how data analysis and PTA configurations can be improved for more sensitive data sets towards the detection of a CGW.

A single GW originating from a SMBHB and passing across the line-of-sight from the radio telescope to the pulsar causes wave-like spacetime distortions at the pulsar's and the Earth's position. Both these distortions, known as the ``Earth term'' (ET) and the ``pulsar term'' (PT), respectively \citep{SesanaVecchio_2010}, affect the measured ToA, causing the GW-induced timing residuals \citep{SesanaVecchio_2010}. The GW frequencies of the ET and PT can differ significantly, depending on the orbital evolution of the SMBHB: due to the GW emission, the SMBHB's orbit shrinks, leading to an increase in the CGW frequency. At higher orbital frequencies, this evolution happens on shorter timescales. For a CGW source emitting at the high-frequency end of the PTA band, $f_\GW > \SI{10}{\nano\hertz}$, and a PTA with a significantly long observation time, the ET and PT frequencies will differ by more than $f_\mathrm{min}$, i.e.\ they fall in different frequency bins and become resolvable \citep{Sesana_2004,SesanaVecchio_2010,IPTA_DR2_CGW}.

The ETs of all pulsars build up coherently \citep{SesanaVecchio_2010}. At the same time, the phase (and for evolving sources, also the period of the PT) depends on the pulsar distance, which for most pulsars is poorly known, with uncertainties typically much larger than the wavelength of the CGWs in question \citep{Deller_2011}. This causes the PT to build up incoherently in the analysis of a PTA data set. This work approaches the PTA configuration aspect by focusing on the role of the PT. Although it has been discussed that a coherent inclusion of the PT in Bayesian PTA analyses is beneficial to improve sky localisation and detection probability \citep{Lee_2011,Arzoumanian_2014, Zhu_2016,Petrov_2025}, the subsequent increase in the search dimension (two parameters, pulsar distance and GW phase, for each pulsar) makes this computationally costly \citep{IPTA_DR2_CGW}. Also, if the ET and PTs are very close, their interference can degrade our ability to observe this signal \citep{Ellis_2012}. In this case, modelling the full signal, including the PT frequencies, (see Eqn.~\eqref{eq:omega_evolution}) requires a precise characterisation of the pulsar distances and the chirp mass of the SMBHB, both of which are often unknown. Thus, in most PTA data analyses \citep[e.g.][]{SesanaVecchio_2010,Petiteau_2013,IPTA_DR2_CGW}, the PTs are treated as an additional source of self-noise and are ignored \citep{Babak_2016,Taylor_2014}. 

As noticed more than a decade ago by \cite{Corbin2010, Ellis_2012}, and investigated in detail by \cite{Zhu_2016}, and several related studies, e.g.\ \cite{Wang_2017,Chen_2022}, using only the ET in a CGW search with unresolved PTs leads to biased results, especially in the source position recovery. Also, the ET-only search is known to detect CGW signatures with a lower S/N than the originally injected signal \citep{Ellis_2012}. In the field, the leading approach to tackle this problem has been increasing the precision of pulsar distance measurements, in order to be able to constrain and efficiently include PTs in the CGW search, \cite[see e.g.][]{Mingarelli_2012,Mingarelli_2018,Kato_2023}.

In this work we aim to explore a different path, namely the behaviour of standard analysis methods in light of resolved PTs. As already pointed out by \cite{Ellis_2012}, it is likely, that analysis methods behave differently in the regime where the ET and PTs do not fall into the same frequency bin. To achieve resolved PTs, we focus on the increase of the difference between the ET frequency and PT frequencies with increasing distance of the pulsars (cf.\ Sec.~\ref{sec:methods}, Eq.~\eqref{eq:time-relation}). Hence we investigate to what extent CGW analysis techniques improve if applied to more distant pulsars with resolved PTs, opposed to an application to nearby pulsars. This targets, on the one hand, the parameter recovery with a Bayesian ET-only search, especially the bias in the source position described by \cite{Zhu_2016}. On the other hand, we investigate the impact of resolved ET and PT frequencies on the Per-Frequency Optimal Statistic \citep{Gersbach_2025}, motivated by the discussion in \cite{Allen_2023}, that the presence of PTs with similar magnitudes and frequencies as the ET, leads to an increased variance of the measured HD curve.

In Section~\ref{sec:bayesian} we focus on the CGW parameter estimation using Bayesian methods, and we investigate to what extent the addition of pulsars at a greater distance to an existing data set, can help improving the CGW parameter recovery. We then present our study on the impact of resolved ET and PT frequencies on the Per-Frequency Optimal Statistic in Section~\ref{sec:PFOS}. We conclude our findings in Section~\ref{sec:conclusion}.


\section{Methods}
\label{sec:methods}

This work explores the analysis of PTA residuals containing a CGW signal using simulated data sets. As we will be exploring in greater detail in the following paragraphs, our analysis setup is as follows: the cornerstone of the simulated residuals are  pulsar ephemerides, which are used to calculate white residuals. On top of these white residuals we add a single CGW signal to all pulsars, in one case assuming all pulsars being located in Earth's proximity, and then setting the pulsars further away, so to achieve a measurable separation of the ET and PT frequency. The resulting data sets are then analysed using both a Bayesian and a frequentist analysis methods. The details of these methods are discussed at the beginning of the respective sections in the paper. 

Throughout this paper we use natural units, where $G = c =1$. For all GW derivations we assume GR. Also, we will use the term `frequency' synonymous with `GW frequency' and refer to any other frequency explicitly.

\subsection{Pulsar timing array data analysis theory}

Pulsar timing array data consist of $N_\mathrm{psr}$ vectors of pulse ToAs, $\Vec{t}_a$, where the subscript $a$ refers to the $a^\mathrm{th}$ pulsar. We will only explicitly use these subscripts for quantities relating to pairs of pulsars. These ToAs are fit using a timing model, which predicts the pulse arrival time by accounting for the pulsar’s rotational behaviour, the astrometry of the pulsar and various other effects, e.g.\ the presence of a binary companion or the dispersion of radio waves due to the ionized interstellar medium, e.g.\ \cite{handbook}. Subtracting the ToAs predicted by the best-fit timing model
leads to the timing residuals $\Vec{\delta t}$. These residuals can be expressed as a sum of multiple contributions, 
\begin{equation}\label{eq:residual_generic}
    \Vec{\delta t} = \mtx{M}\Vec{\epsilon} + \Vec{n} + \Vec{s},
\end{equation}
where $\mtx{M}$ is the design matrix describing the linearised timing model, $\Vec{\epsilon}$ is the vector of the timing model parameter errors, $\Vec{s}$ can be any additional deterministic signal, such as a CGW, and $\Vec{n}$ refer to the stochastic signals present in the residuals, also called the pulsar noise budget. 

The noise present in all residuals leads to correlations between individual residuals. The noise processes are modelled as a sum of Gaussian Processes, characterised via their power spectral densities (PSDs) \citep{vanHaasterenLevin_2013}. The majority of the PSDs is modelled as power laws, $P_X(f) \sim A_Xf^{-\gamma_X}$, where $X$ refers to the name of the signal, and $A_X$ and $\gamma_X$ are its amplitude and spectral index. 
The noise budget of a single pulsar has three major contributions: (i) the white noise (WN, flat PSD), arising from noise in the instrumentation and characterised per radio telescope backend with the WN parameters EFAC, $\mathcal{I}$, and EQUAD, $\mathcal{Q}$, and (ii) red noise (RN) processes ($\gamma_\mathrm{RN} > 0$), distinguished into (iia) achromatic red noise (amplitude does not depend on the observational frequency), or pulsar timing noise, caused by random changes in the pulsar's spin frequency and (iib) chromatic red noise (amplitude depends on the observational frequency as $A \propto f_\mathrm{obs}^{-\gamma}$, with $\gamma > 0$), introduced by variations in the interstellar medium (ISM) \citep{Lentati2014}. 

A stochastic GW signal present in the PTA data set gives rise to additional noise that is correlated between pulsar \citep{AllenRomano_1999}, i.e.\ it can be detected by investigating the cross-correlations between residuals of different pulsars. The cross-correlation matrix $S_{ab}$ is given as
\begin{align}\label{eq:cc-matrix}
    \mtx{S}_{ab} &= \langle\Vec{\delta t}_a\Vec{\delta t}_b^T\rangle_{a\neq b} \\
     &= \Vec{F}_a \; \Gamma_{ab} P_\GWB(f) \; \Vec{F}_b^T, 
\end{align}
with the HD correlation coefficients $\Gamma_{ab}$ and the power spectral density of the gravitational-wave signal
\begin{equation}\label{eq:GWB-PSD}
    P_\GWB(f) = \frac{A_\GWB^2}{12\pi^2}\left(\frac{f}{f_\mathrm{yr}}\right)^{-\gamma_\GWB} ,
\end{equation}
which we again parametrize as a power law in terms of an amplitude, $A_\GWB$, and a spectral index, $\gamma_\GWB$. In the narrow-band analysis of the power spectral density we investigate in Sec.~\ref{sec:PFOS}, the HD correlation will only be evaluated in a single frequency bin, not across the full spectrum.

\subsection{Continuous gravitational wave model}
\label{ssec:cgwmodel}
Assuming a single CGW signal is present in the PTA data set, the additional deterministic contribution $\Vec{s}$ in Eq.~\eqref{eq:residual_generic} is variation caused by the CGW, which can be derived from GR. The strain of a CGW emitted by a circular SMBHB at right ascension $\phi_\GW$ and co-latitude $\theta_\GW$ induces the residuals 
\begin{equation}\label{eq:CGW_residuals}
    s_a(t,\Hat{\Omega})  = \sum_{A = +,\cross} F^A_a(\Hat{\Omega}) \cdot \left[s_A(t-\tau_a) - s_A(t)\right]
\end{equation}
in the data set of the $a^\mathrm{th}$ pulsar of a PTA. $s_A(t)$ denotes the ET and $s_A(t-t_a)$ the PT. 
The $F^A$ are the antenna pattern functions \citep{Gair_2014},
\begin{align}
    F_a^+(\Hat{\Omega}) &= \frac{1}{2} \frac{(\Hat{m}\cdot\Hat{p}_a)^2 - (\Hat{l}\cdot\Hat{p}_a)^2}{1-\Hat{\Omega}\cdot\Hat{p}_a}, \\
    F_a^\times(\Hat{\Omega}) &= \frac{(\Hat{m}\cdot\Hat{p}_a)(\Hat{l}\cdot\Hat{p}_a)}{1-\Hat{\Omega}\cdot\Hat{p}_a},
\end{align}
expressed in terms of the polarisation vectors $\Hat{l}$ and $\Hat{m}$ \citep{SesanaVecchio_2010}.

The delay time, $t_a$, between the source and the $a^\mathrm{th}$ pulsar can be evaluated geometrically from the distance between the Earth and the pulsar, $d_a$ as
\begin{equation}\label{eq:time-relation} 
    \tau_a = d_a(1\mathbf{-}\Hat{\Omega}\cdot\Hat{p}_a),
\end{equation}
where $\Hat{p}_a$ is the unit vector pointing from the Solar System Barycenter to the pulsar $a$, and $\Hat{\Omega}$ denotes the unit vector pointing opposite to the CGW propagation direction, from the Earth to the GW source.

The wave forms are given as
\begin{align}
    s_+(t) = \frac{h_0}{\omega(t)} &\left[ \left(1+\cos^2\iota \right) \cos{2\psi} \sin(\Phi(t)) \right. \\
    & \left. +2\cos\iota \sin{2\psi} \cos(\Phi(t)) \right],\\
    s_\times(t) = \frac{h_0}{\omega(t)}&\left[\left(1+\cos^2\iota \right) \cos{2\psi} \cos(\Phi(t)) \right. \\
    &\left. -2\cos\iota \sin{2\psi} \sin(\Phi(t)) \right]
\end{align}
with the GW frequency, $\omega(t) = 2\pi f_\GW(t)$, the GW phase, $\Phi(t)$, as well as the inclination of the binary, $\iota$, and the GW polarisation angle $\psi$. The intrinsic strain amplitude, $h_0$, is defined as
\begin{equation}
    h_0 = 2\frac{\mc^{5/3}(\pi f_\GW)^{2/3}}{d_L},
\end{equation}
in terms of the chirp mass, $\mc$, and the luminosity distance to the source, $d_L$, \citep{Zhu_2016}.

In this work we consider only slowly evolving binaries, meaning that we consider their frequency evolution over the travel time of the radio pulse from emission at the pulsar until its reception on Earth, but we assume that the difference between ET and PT frequencies gained is constant over the PTA observation time span. In this approximation, the ET and PT frequencies to leading order of the radiation reaction equation are given as
\begin{align}
    \omega_\mathrm{E} &= \omega_0 = 2\pi f_\GW, \\
    \omega_a &= \omega_0 \left[1- \frac{256}{5}\mc^{5/3} \omega_0^{8/3} \cdot d_a (1-\Hat{\Omega}\cdot\Hat{p}_a)\right],\label{eq:omega_evolution}
\end{align}
meaning that the ET always has a higher frequency than the PT. The corresponding phases are given as
\begin{align}
    \Phi_\mathrm{E}(t) &= \Phi_0 + \omega_0 t, \\
    \Phi_\mathrm{a}(t) &= \Phi_0 + \Phi_at + \frac{1}{32\mc^{-5/3}} \left(\omega_0^{-5/3} - \omega_a^{-5/3}\right), \label{eq:pt_phase}
\end{align}
for a full derivation \cite[see e.g.][]{SesanaVecchio_2010}.

If the source has significantly evolved over the pulse travel time, the ET and PTs frequencies can differ more than the fundamental frequency of the PTA, such that they fall into different frequency bins. This is usually the case for GW frequencies above \SI{10}{\nano\hertz}. Due to the limited frequency evolution at lower frequencies, both terms will have the same frequency within the resolvability of the PTA for CGWs with an ET frequency below $\sim\SI{10}{\nano\hertz}$, for current PTA time spans.

\subsection{Simulation details}
\label{ssec:simulation}

For this work we use realistic data set simulations based on the EPTA DR2 data set \citep{EPTA_DR2_DATA}. We assume the ToA uncertainties of the DR2new data set, together with the observational time span of $\sim\!10$ years and a mean cadence of the pulsars based on the real PTA data set.
All our simulations are performed using the \textsc{python} package \textsc{libstempo}\footnote{\url{https://github.com/vallis/libstempo}}.

\begin{figure}
    \centering
    \includegraphics[width=\linewidth]{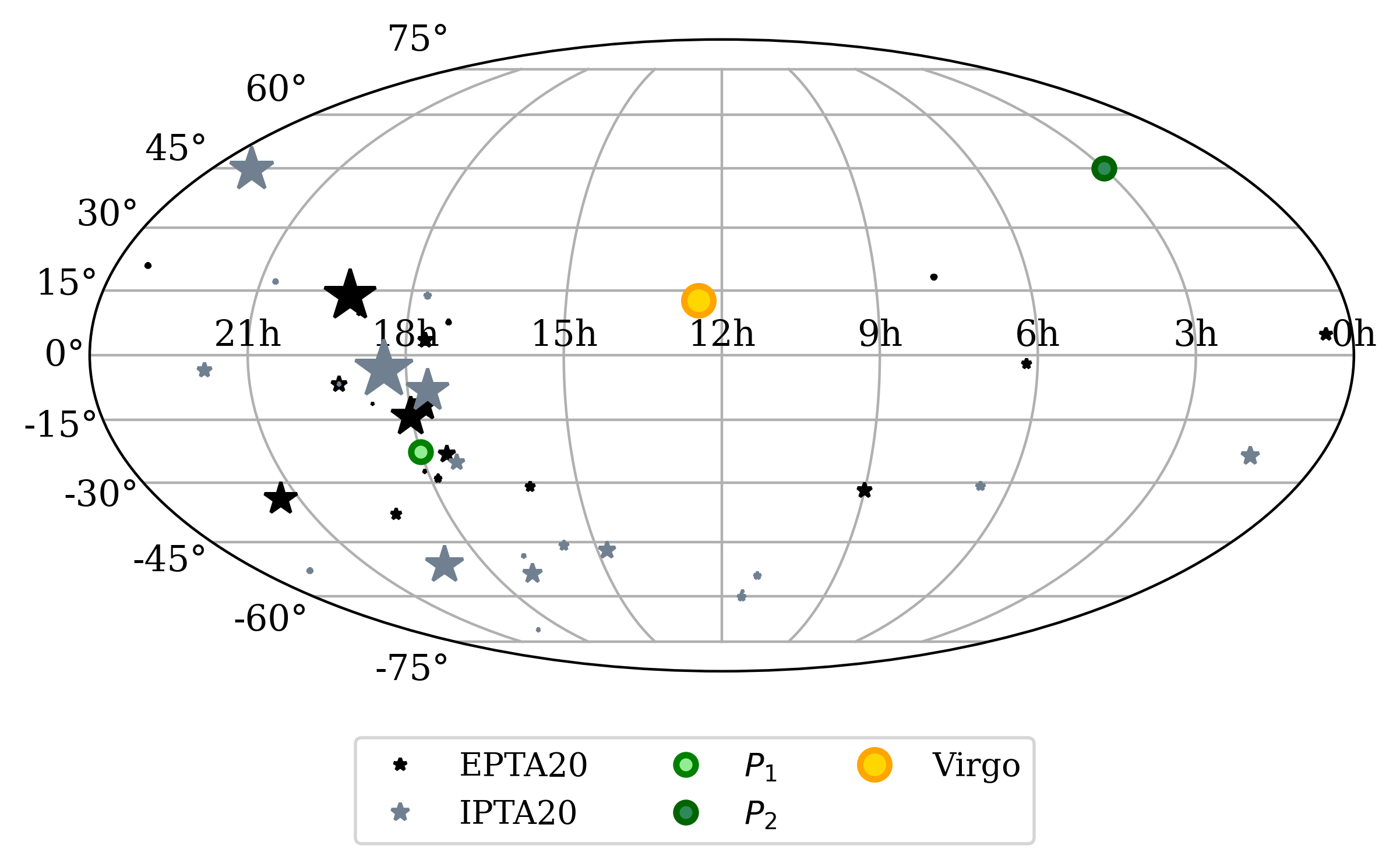}
    \caption{Sky map indicating the positions of the pulsars (stars) used in this work, and the CGW positions used for the investigations in this wok. The size of the pulsar position markers is inverse proportional to the residual root-mean-square of each pulsar.}
    \label{fig:skymap_pulsars_cgws}
\end{figure}

The CGW signal injected to the ToAs is calculated according to Eq.~\eqref{eq:CGW_residuals}, setting $\iota = \pi$, $\psi = \pi/2$, $\Phi_0 = \pi/2$. The distribution of PTs in a PTA data set is not only dependent on the pulsar distance, but also on the angular separation between the pulsar and the CGW source. By default, we choose to inject the CGW at the sky position of the Virgo cluster ($\alpha = \SI{3.2594}{\radian}, \delta=\SI{0.2219}{\radian}$), because it is proposed, \cite[e.g.][]{Simon_2014}, that it is the most probable host of a SMBHB that would be visible as a single CGW source. In order to test the dependence of the effects investigated in this work on the pulsar-source angular distance, we also run selected analyses on an injected CGW source at two standout positions, shown as the green circles in Fig.~\ref{fig:skymap_pulsars_cgws} and described in more detail in Sec.~\ref{ssec:PTA_setup_and_source_position}.

The remaining CGW parameters are chosen considering (a) that the combination of $f_\gw$ and $\lmc$ allows for such a frequency evolution, that the transition from unresolved PTs to resolved PTs is realisable using reasonable pulsar distances, and (b) that the S/N of the injected signal, following the definition from \cite{Zhu_2016},
\begin{equation}
    \rho^2 = \sum_{j = 1}^{N_\mathrm{PSR}} \rho_j^2 = \sum_{j = 1}^{N_\mathrm{PSR}} \sum_{i = 1}^{N_\mathrm{ToA}} \left[\frac{s(t_i)}{\sigma_j}\right],
\end{equation}
falls close to the intermediate ($\rho\sim30$) signal regime, where $s(t_i)$ is the injected CGW signal and $\sigma_j$ is the mean ToA error of the $j^\mathrm{th}$ pulsar. As pointed out by \cite{Zhu_2016}, the strong S/N regime has been ruled out for the Virgo cluster by recent PTA analyses, nonetheless it is helpful to test the method.

\begin{figure}
    \centering
    \includegraphics[width=\linewidth]{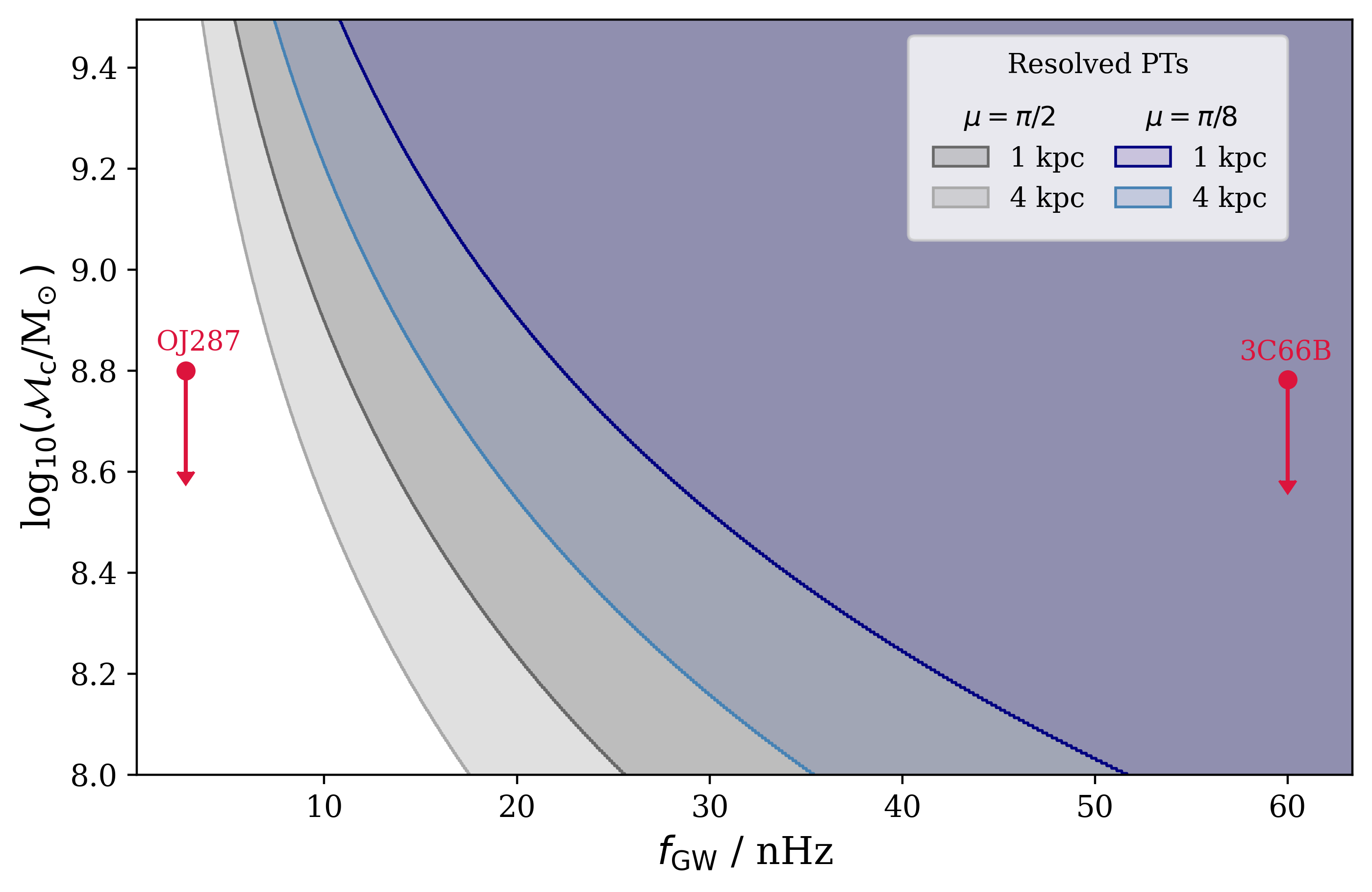}
    \caption{$f_\GW$-$\lmc$ p$f_\GW$-$\lmc$ parameter space. The shaded areas indicate the region of resolved PTs, for a PTA timing baseline of 10 years. The gray areas correspond to an angular separation between the pulsar and the CGW source of \SI{90}{\degree}, the blue areas correspond to \SI{22.5}{\degree}. The darker and lighter shades refer to a pulsar distance of \SI{1}{\kilo\parsec} and \SI{4}{\kilo\parsec}, respectively. The red symbols indicate the positions of two exemplary SMBHB candidates in that parameter space, with $f_\GW$ estimates and upper limits of $\lmc$ from electromagnetic models or PTA analyses (OJ287: \cite{Titarchuk_2023,Komossa_2023}, 3C66B: \cite{CardinalTremblay_2025}).}
    \label{fig:flmc_parameterspace}
\end{figure}

We find that for the EPTA data set, these conditions are met for $\lmc=9.1$ and $f_\GW=\SI{15.9}{\nano\hertz}$, corresponding to the sixth frequency bin of the PTA, if we compare pulsars at \SI{1}{\kilo\parsec} and \SI{4}{\kilo\parsec} distance. This set up is our fiducial choice of parameters, later: Case (1), to which we will compare other set ups in order to illustrate other parameter dependences besides the pulsar distance. The $f_\GW-\lmc$ parameter space is visualised in Fig.~\ref{fig:flmc_parameterspace}, together with the regions for which the PT is resolved, given a timing baseline of \SI{10}{\yr} and two different values of pulsar-CGW angular separation, $\mu$. Evidently, the $f_\GW$-range, in which the transition from unresolved to resolved PTs can be tested, is reasonably narrow, illustrating our specific choices of $\lmc$ and $f_\GW$.

To aid comparability across all our simulations, we henceforth fix the scheme of ``nearby pulsars'' sitting at \SI{1}{\kilo\parsec} distance, and ``distant pulsars'' being placed at \SI{4}{\kilo\parsec} distance. This leaves the parameters $d_L$ and $\lmc$ to regulate the frequency evolution, strain amplitude and $S/N$ of the injected signal. 

We assume that the pulsars at $\SI{4}{\kilo\parsec}$ have similar ToA uncertainties as the pulsars at $\SI{1}{\kilo\parsec}$. Although this is counter-intuitive following the well-known scaling between the ToA uncertainty and S/N, which is intimately linked to the pulsar distance, we argue, that in light of the FAST radio telescope or the Square Kilometre Array, it is reasonable to assume that pulsars at moderately larger distances (such as $\SI{4}{\kilo\parsec}$) will yield similar ToA uncertainties as those obtained from observations with current radio telescopes. With the larger observational frequency bandwidths and better telescope gains of these next-generation facilities, the timing precision of most PTA sources will be jitter- rather than sensitivity-limited. For example, this can be achieved by adapting the integration time, i.e.\ weaker pulsars will be observed longer, a strategy that is for instance used in the current scheduling of the MeerKAT PTA \citep{Miles_2022,Middleton2025}.

To explore different regimes of CGW strain, and to constrain the effect of the PT resolution, we analyse a total of three different CGW injection cases:
\newline
\newline
\hspace*{1em}
\begin{tabular}{p{1.7cm}p{6cm}}
$-\;$ Case (1): &resolved PTs for $d_\mathrm{p}=\SI{4}{\kilo\parsec}$;  \\[0.2em]
                &$\quad\lmc=9.1$; $d_L = \SI{15}{\mega\parsec}$; \\[0.2em]
                &$\quad h_{0,(1)} = \num{9.9e-15}$; S/N $\sim 36$\\[0.5em]
$-\;$ Case (2): &resolved PTs for $d_\mathrm{p}=\{\SI{1}{\kilo\parsec}$, $\SI{4}{\kilo\parsec}\}$; \\[0.2em]
                &$\quad\lmc=9.5$; $d_L = \SI{70}{\mega\parsec}$;\\[0.2em]
                &$\quad h_{0,(2)} = h_{0,(1)}$; S/N $\sim 36$ \\[0.5em]
$-\;$ Case (3): &resolved PTs for $d_\mathrm{p}=\SI{4}{\kilo\parsec}$,  \\[0.2em]
                &$\quad\lmc=9.1$; $d_L = \SI{3}{\mega\parsec}$;\\[0.2em]
                &$\quad h_{0,(3)} = 5 h_{0,(1)}$; S/N $\sim 181$.\\[1em]
\end{tabular} 

\noindent
The particular choices of $d_L$ and $\lmc$ enable the following tests: By comparing Cases (1) and (2), we investigate whether changing to resolved PTs yields a singular improvement of the CGW analysis method or if any improvement increases with a larger frequency difference between ET and PTs -- while maintaining the same S/N in both cases. On the other hand, comparing Cases (1) and (3) will demonstrate the impact that a larger S/N has on any improvement effect -- while maintaining the same frequency evolution of the source.

We note, that due to the angular proximity to the fiducial CGW position of some pulsars in the EPTA data set, even at \SI{4}{\kilo\parsec}, their PTs are not resolved. In order to single out the effect of resolved PT, we restrict ourselves to use only 20 out of the 25 EPTA pulsars, for which the PTs are fully resolved at a mean pulsar distance of \SI{4}{\kilo\parsec}. \cite{Speri_2023} demonstrated that for both the latest IPTA data set (IPTA DR2) and EPTA data set (EPTA DR2), $\sim95\%$ of the recovered S/N of the GW signal can be recovered using the most responsive $\sim 20$ pulsars of these data sets, so assuming a 20-pulsar PTA is sufficiently realistic. We will refer to this subset as the \textsc{EPTA20} data set. The resulting pulsar sky distribution is shown in Fig.~\ref{fig:skymap_pulsars_cgws}, where the size of the markers is inverse proportional to the residual root-mean-square of each pulsar.

Similar to \cite{Zhu_2016}, we analyse 500 realisations of the data sets. The analysis methods relevant for the respective section are described at the beginning of each section.

\section{Bayesian CGW search}
\label{sec:bayesian}

It is common amongst the PTA community to constrain the posterior distributions of PTA parameters in a Bayesian approach by sampling the PTA likelihood function \citep{vanHaasteren2009,vanHaasterenLevin_2013}
\begin{equation}\label{eq:likelihood}
    \mathcal{L} = \frac{1}{\sqrt{\det 2\pi \mtx{C}}}\exp\left\{ \Vec{\delta t}^T \mtx{C}^{-1} \Vec{\delta t}^T \right\}
\end{equation}

In sampling the PTA likelihood function, Eq.~\eqref{eq:likelihood}, the posterior distribution of the CGW parameters $\log_{10}f_\GW$, $\lmc$, $\log_{10}h$, $\phi_\GW$, $\cos\theta_\GW$, $\cos\iota$, $\Phi(t)$, and $\psi$ are determined. If the PT is not considered in the model, the chirp mass cannot be constrained. 

We will employ the most commonly used analysis scheme used in the PTA community, \citep[e.g.][]{IPTA_DR2_CGW,EPTA_DR2_GWB,NG15_GWB,PPTA_GWB}, namely a Markov Chain Monte Carlo (MCMC) sampler, \textsc{PTMCMC} \citep{ptmcmc}, and the \textsc{enterprise} software \citep{enterprise} to build the PTA model and calculate the likelihood.

\subsection{Test case scenario results}

\begin{figure}[htbp]
    \centering
    \subfigure[Case (1) with white and red noise, $\lmc=9.1, d_L=\SI{15}{\mega\parsec}$]{\includegraphics[width=\linewidth]{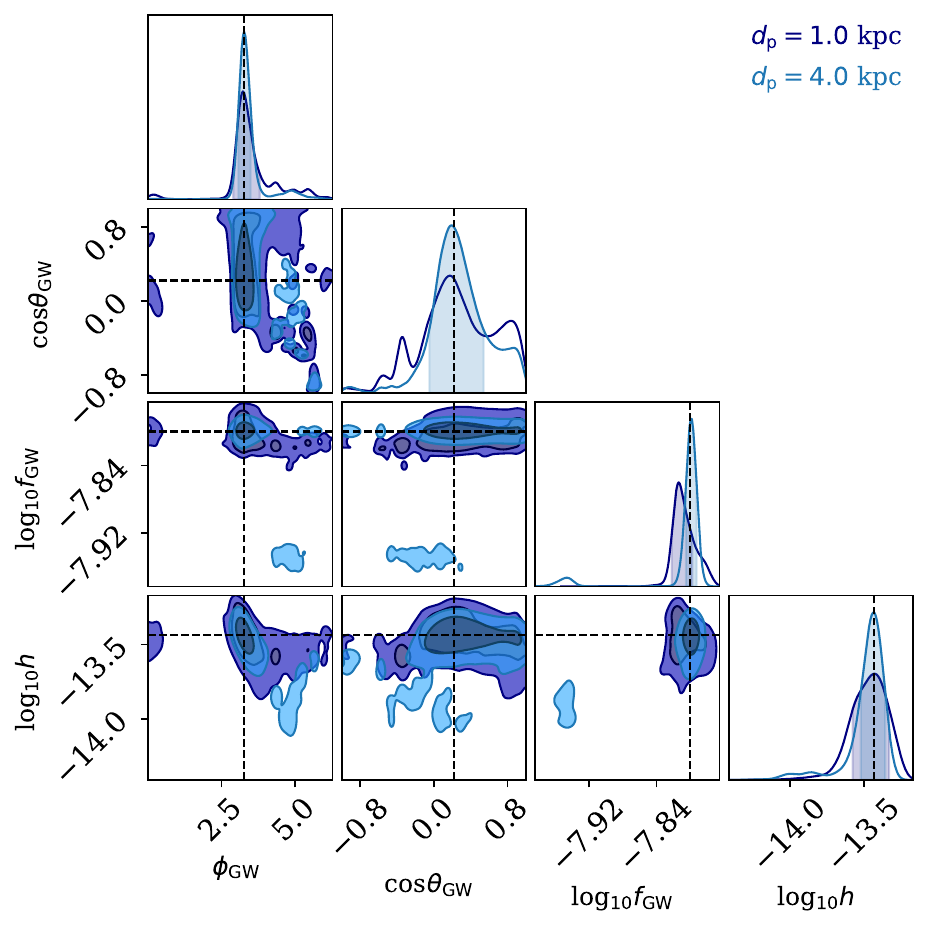}}
    \subfigure[Case (1) with white noise only, $\lmc=9.1, d_L=\SI{15}{\mega\parsec}$]{\includegraphics[width=\linewidth]{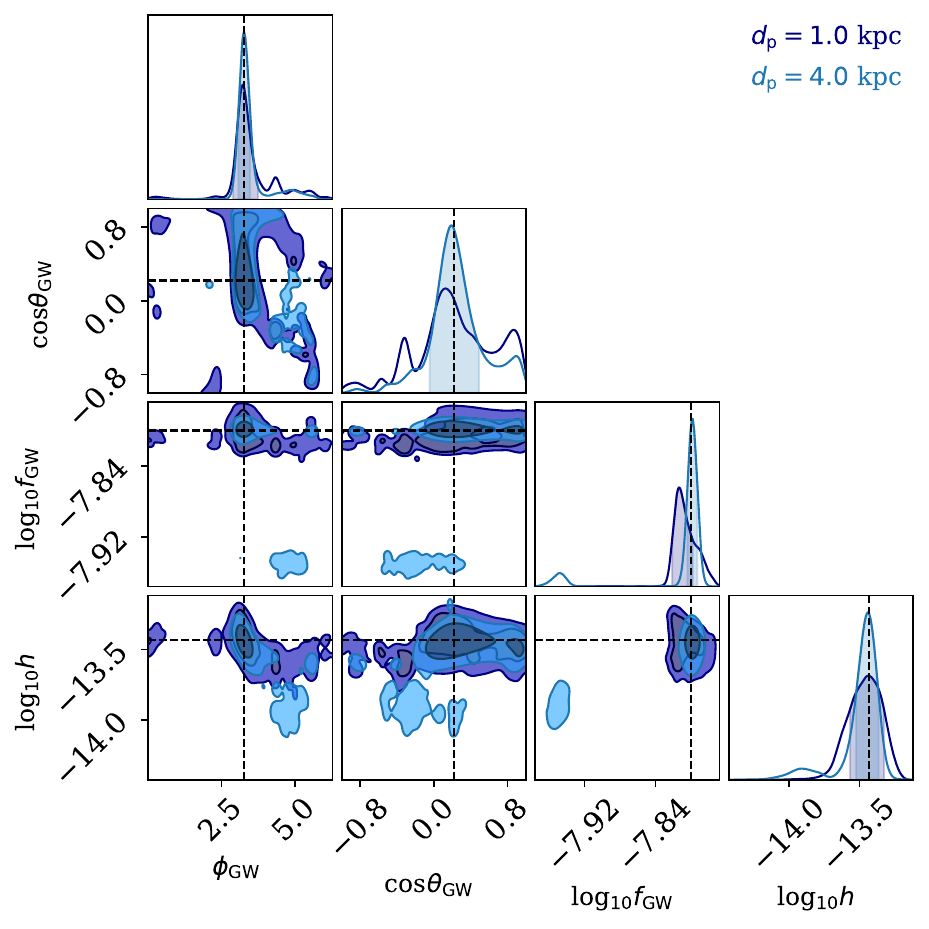}}
    \caption{Cornerplot of the CGW parameter posteriors (excluding $\lmc$, $\Phi_0$, $\psi$) created from the joint MCMC chain of 500 realisations of each data set. It shows the result from the data set containing a CGW signal with $\lmc=9.1$. The darker contours correspond to the PTA with pulsars at \SI{1}{\kilo\parsec}, the lighter contours correspond to the PTA with pulsars at \SI{4}{\kilo\parsec}}
    \label{fig:epta20-WN-RN-CGW}
\end{figure}

In this part we aim to investigate, if resolved PTs can lead to improved posteriors of an ET-only MCMC search, compared unresolved PTs. In order to keep the PTA data set as realistic as possible, we analyse the \textsc{EPTA20} data set as described above, once with pulsars at \SI{1}{\kilo\parsec} and once with pulsars at \SI{4}{\kilo\parsec}. Real PTA data sets are not only affected by WN, but also by RN. As RN processes are known to be covariant with GW parameters, we test the influence of additional RN present in the data set on the outcome of our simulation. Thus, we start by analysing the Case (1) simulation, once in a data set with WN only, and once in a data set that also has RN present in each pulsar ToA series. For the simulations, we randomly choose the RN amplitude and spectral index for each pulsar from $\log_{10}A_\mathrm{RN}\in[-15.5,-13.5]$ and $\gamma\in[1.7,2.2]$. The resulting cornerplots from analysing 500 realisations of each setup is shown in Fig.~\ref{fig:epta20-WN-RN-CGW}.

First, we find, that in both analyses, the marginalised CGW parameter posteriors obtained from the \SI{4}{\kilo\parsec}-PTAs (i.e.\ with resolved PTs) are either narrower and/or more accurate. Not only for the source position parameter distributions, $\cos\theta_\GW$ and $\phi_\GW$, but also the posteriors of the GW strain, $\log_{10}h$, and the GW frequency, $\log_{10}f_\GW$.

The posterior of the GW frequency exhibits an interesting behaviour for the \SI{4}{\kilo\parsec}-setup. Next to the dominant peak, exactly at the ET frequency, it features a second, smaller peak at a slightly lower frequency. Comparing its position to the distribution of PT frequencies in the PTA setup, we find that it coincides very well. So we assume that this additional peak is likely caused by an occasional confusion of the ET frequency with the PT frequency during the MCMC sampling. The ET-only search tries to fit a single sinusoidal pattern with a frequency commonly found in all pulsars, which predominantly is the case at the ET frequency. But also, do to the random noise fluctuations present in all pulsars, there can be some correlation picked up by chance at the PT frequencies. This apparent correlation is only a coincidence, unlike the actually correlated signal at the ET frequency. Nonetheless, it causes the MCMC chain to stray into that area of the parameter space. Hence the second peak is formed, but the peak around the ET is more significant. At the same time, this allows us to understand why the same posterior of the \SI{1}{\kilo\parsec}-setup is broader at the position of the injected CGW frequency. Here, the PTs and the ET are not resolved, hence the occasional confusion of the MCMC chain into the PT frequencies becomes visible as a broadened posterior distribution.

Additionally we have demonstrated, that the WN-only simulation produces representative results of the more realistic WN+RN simulation. This is not surprising, since with $f_\GW =\SI{15.9}{\nano\hertz}$, we are dealing with the higher-frequency regime of a PTA data set, at which the influence of RN is diminished. So in order to reduce complexity of the setup and single out the PT effect, we henceforth restrict ourselves to the more simplistic WN-only simulations.

\begin{figure}[htbp]
    \centering
    \subfigure[Case (2) with WN, $\lmc=9.5, d_L=\SI{70}{\mega\parsec}$]{\includegraphics[width=\linewidth]{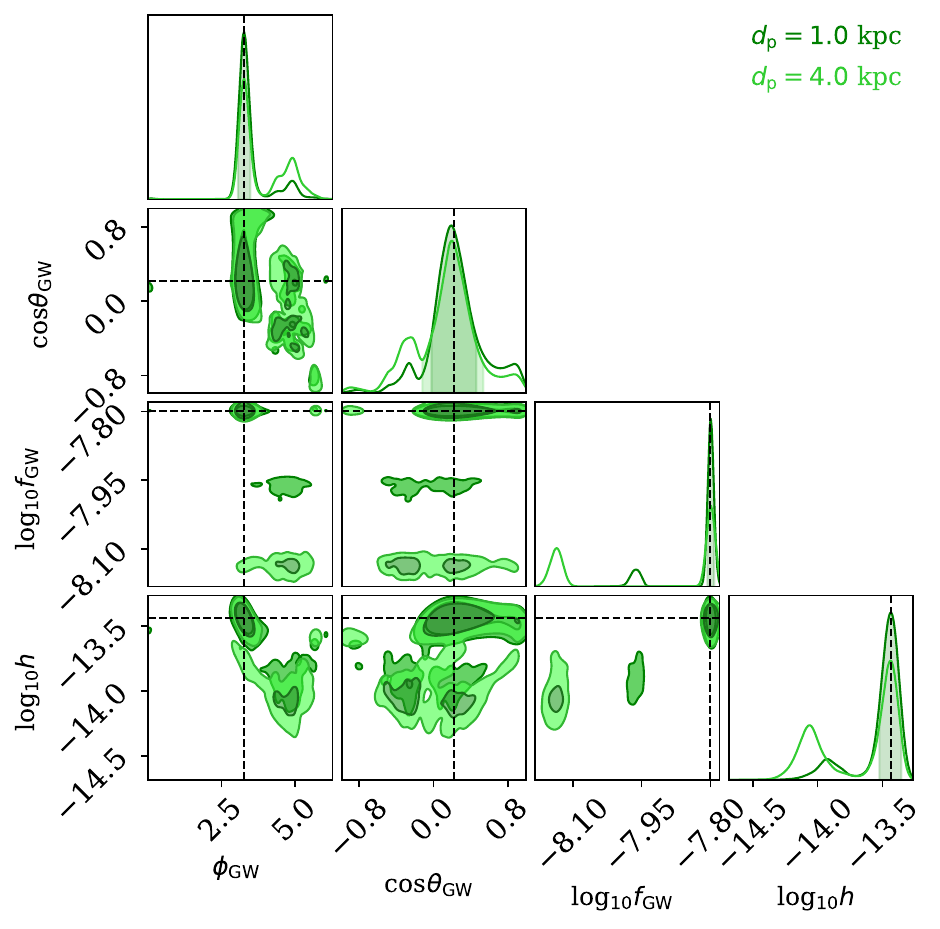}}
    \subfigure[Case (3) with WN, $\lmc=9.1, d_L=\SI{3}{\mega\parsec}$]{\includegraphics[width=\linewidth]{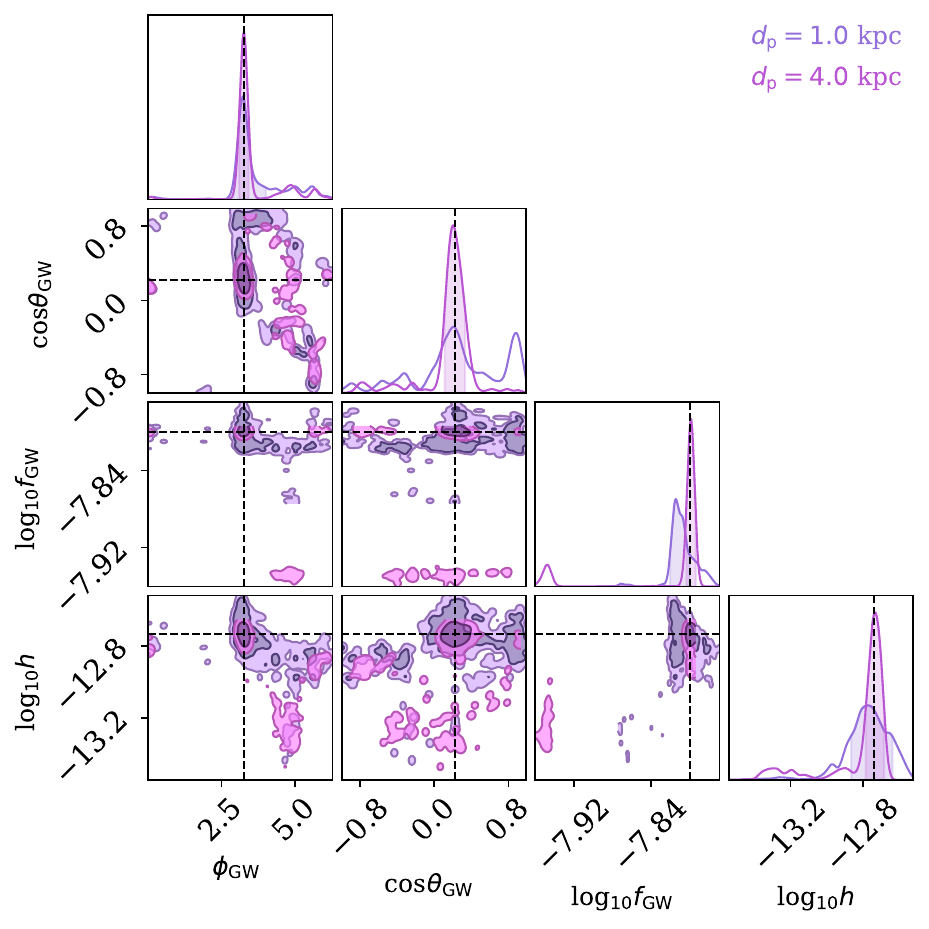}}
    \caption{Cornerplot of the CGW parameter posteriors (excluding $\lmc$, $\Phi_0$, $\psi$) created from the joint MCMC chain of 500 realisations of each data set. The darker contours correspond to the PTA with pulsars at \SI{1}{\kilo\parsec}, the lighter contours correspond to the PTA with pulsars at \SI{4}{\kilo\parsec}}
    \label{fig:epta20-case2and3}
\end{figure}

In the next step, we look into the behaviour of this effect using the Case (2) and Case (3) CGW injections, again for the \textsc{EPTA20} pulsars at either \SI{1}{\kilo\parsec} or \SI{4}{\kilo\parsec}. The resulting cornerplots of the CGW parameter posteriors are shown in Fig.~\ref{fig:epta20-case2and3}.

In Case (2), the PTs are resolved for both pulsar distances. Clearly, the marginalised posteriors of the individual parameters do not exhibit any significant improvements at a larger distance. The only notable difference in the posteriors of Case (2) is found in the distribution of $\log_{10}f_\GW$. This shows, that analysis performance cannot be amplified by a PT separation of multiple frequency bins compared to a small separation by a single frequency bin. Hence we conclude, that for the Bayesian analysis, there is only a single improvement step, caused by the transition from unresolved to resolved PTs. 

With Case (3) we demonstrate that the improvement achieved with larger pulsar distances is similarly visible for an increased source strain. Notably, the secondary peak in the posterior of $\log_{10}f_\GW$ became slightly larger. This is not surprising, as the larger signal strength amplifies the side effect that the MCMC chain occasionally samples the PT frequencies.

Overall this demonstrates that the observed enhancement effect is caused by the transition from unresolved to resolved PTs.

\subsection{PTA setup and source position}
\label{ssec:PTA_setup_and_source_position}

In general, whether it is possible to constrain the parameters of a CGW signal present in a PTA data set depends on a wide range of characteristic properties of the PTA data set, such as the number of pulsars, their noise properties and the overall S/N of the signal as well as the pulsars' individual S/N contribution.

We have demonstrated, that with sufficient S/N of the signal, resolved PTs can improve the parameter recovery of a Bayesian ET-only CGW analysis. There are two main factors setting a natural limit to this effect -- the distribution of pulsars on the sky, in combination with their individual S/N contribution to the CGW detection in the PTA. As it can be seen in Eq.~\eqref{eq:omega_evolution}, the PTs' frequency difference does not only depend on the pulsars' distance, but also on their location with respect to the CGW source. If the pulsars are too close to the source, $1 - \Hat{\Omega}\cdot\hat{p}_a\sim0$, and so $\omega_a \sim \omega_\mathrm{E}$, even for large $d_a$. In this case, even pulsars at large distances have unresolved PTs. On the other hand, if the source is located at a greater angular distance to some of the pulsars, but their S/N contribution is not significant, there is no visible improvement of the posteriors, as its behaviour is dominated by the higher S/N pulsars. 

\begin{figure}
    \centering
    \includegraphics[width=\linewidth]{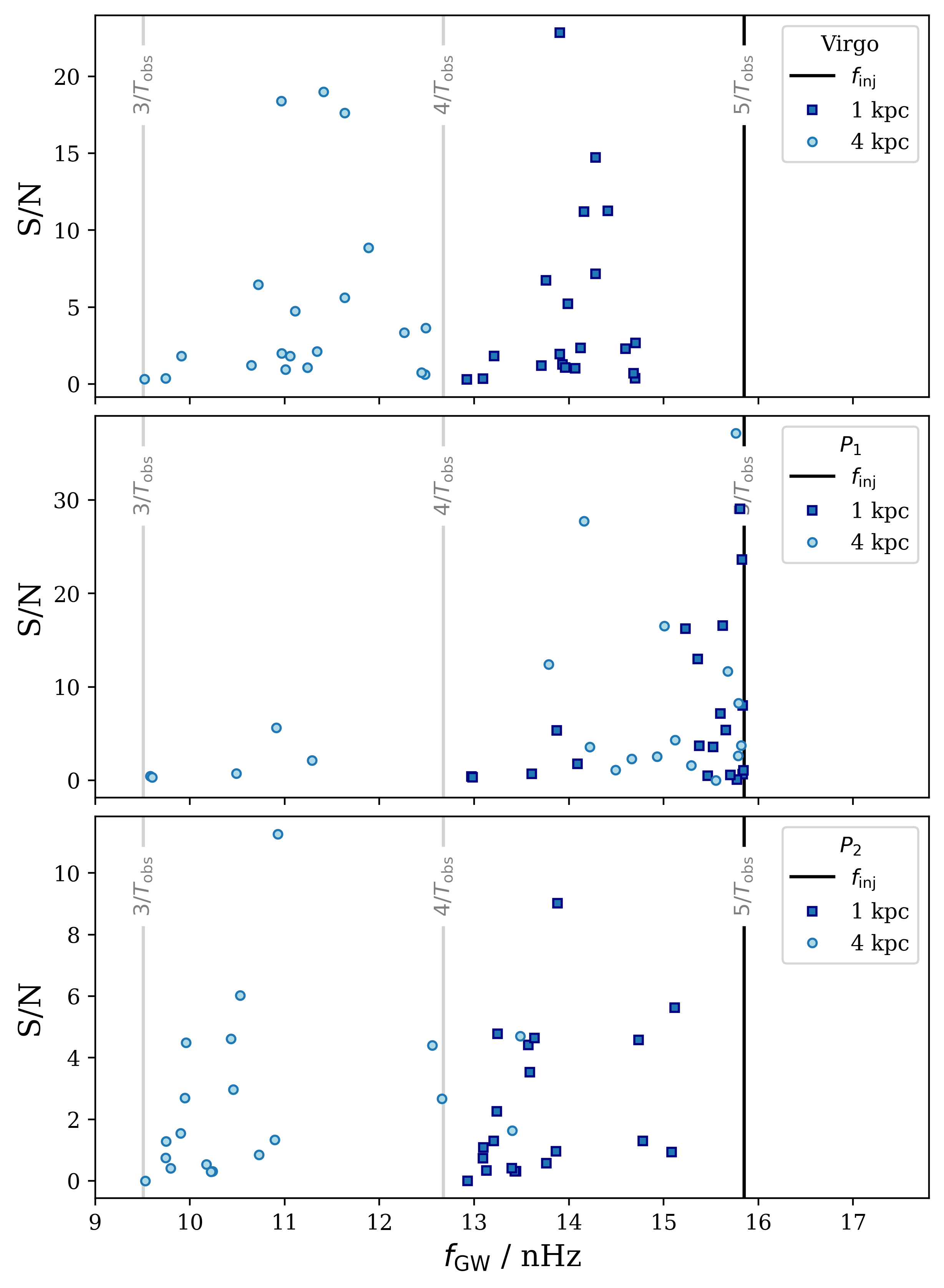}
    \caption{Distribution of PT frequencies versus S/N contribution in the \textsc{EPTA20} data set for three different CGW source positions and pulsars at both \SI{1}{\kilo\parsec} (dark blue squares) and \SI{4}{\kilo\parsec} (light blue dots). The CGW frequency (ET frequency) is indicated as the dark blue vertical line. The fundamental frequencies of the PTA with $T_\mathrm{obs}=\SI{10}{\yr}$ are indicated with the gray vertical lines and labels. Top: Virgo cluster. Middle: $P_1 = $(RA 18h, DEC\SI{-22.5}{deg}). Bottom: $P_2 = $(RA 3h, DEC\SI{-45}{deg}). Due to the dependence of the strain on the angular distance between the pulsar and the source (cf.\ Eq.~\eqref{eq:CGW_residuals}), the $S/N$ varies across the three positions.}
    \label{fig:pt_vs_sn_EPTA20-CGWpos}
\end{figure}

We have tested these limiting cases by changing the source position from the Virgo cluster to two other positions, $P_1 =$ (RA 18h, DEC\SI{-22.5}{deg}) and $P_2 =$ (RA 3h, DEC\SI{-45}{deg}). $P_1$ resembles a location in the vicinity of the majority of pulsars, whereas $P_2$ is at a larger distance to pulsars with a large RMS. For each of the three cases, the distribution of the PT frequency versus the S/N of the signal in each pulsar is shown in Fig.~\ref{fig:pt_vs_sn_EPTA20-CGWpos}. For these calculations we have used the Case (1) set up.

Evidently, the improvement will be highly dependent on the pulsar constellation. While for a CGW at (and around) $P_1$, the S/N of the signal is high, promising a certain detection, only a few pulsars have resolved PTs at \SI{4}{\kilo\parsec}, and those carry only a marginal fraction of the S/N. For $P_2$, most of the PTs are resolved at the larger pulsar distance, but the overall S/N of the signal is very low, due to the change in the angular distance between the pulsars and the CGW source, cf.\ Eq.~\eqref{eq:CGW_residuals}. Hence in this case, we can expect that the improvement is significantly impaired by the poor detectability of the signal, and hence will likely not be evident. In comparison, we find for the source position close to the Virgo cluster sky position, that most of the PTs are resolved at \SI{4}{\kilo\parsec}, and that the S/N is sufficiently large. Thus, the effect is visible.

\subsection{Adding pulsars to an existing PTA}

\begin{figure}[htbp]
    \centering
    \includegraphics[width=\linewidth]{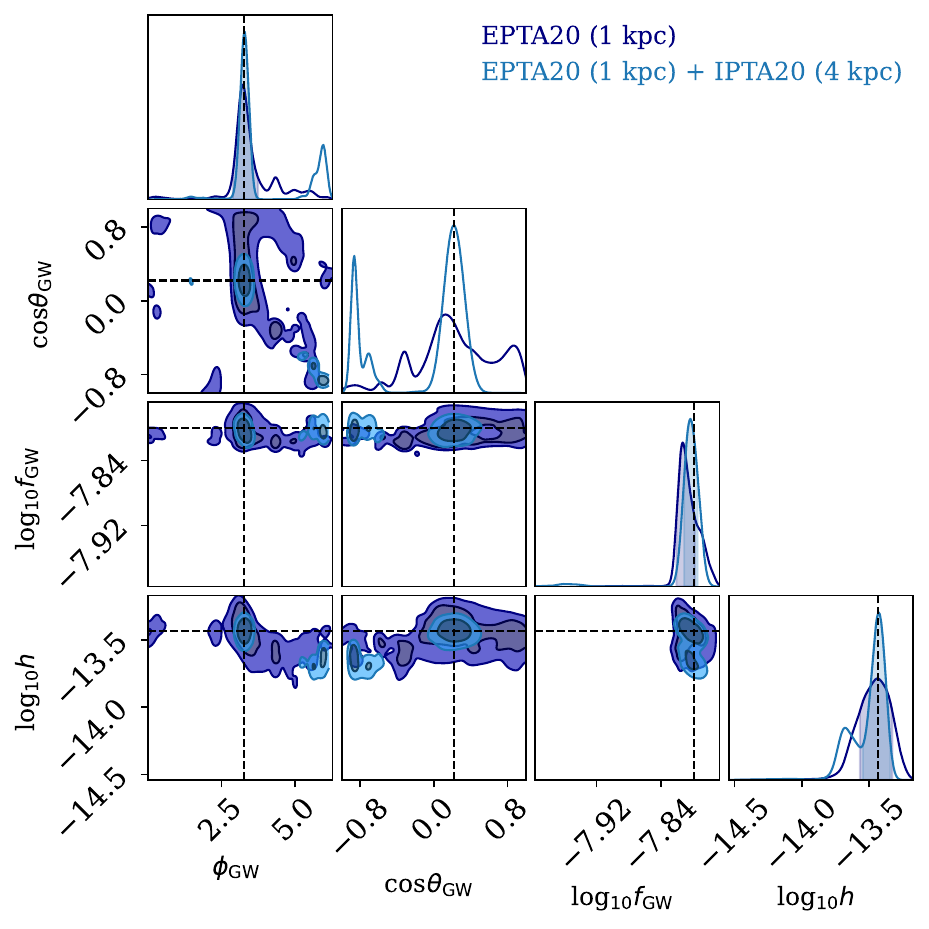}
    \caption{Bayesian posteriors for 500 realisations of the Case (1) injection. The darker contours correspond to the \textsc{EPTA20} data set, the lighter contours to the \textsc{EPTA20+IPTA20} data set, that has a S/N increase of 40\%.}
    \label{fig:epta-addon}
\end{figure}

The way we investigated the behaviour of the Bayesian analysis method in the previous subsections corresponds to analysing only a subset of more distant pulsars in a PTA data set. Actual PTA data sets span multiple decades, and also some of the most precisely timed pulsars are at a comparably small distance to Earth. Moreover, at the current state, PTA datasets only contain a limited number of pulsars, so it is not feasible to form meaningful subsets in the near future. Hence it is more realistic to investigate the effect that adding pulsars at a greater distance has on the Bayesian parameter recovery of a CGW signal. 
We test this by adding 20 pulsars to the \textsc{EPTA20} data set, these are shown as gray stars in Fig.~\ref{fig:skymap_pulsars_cgws}, with the size of the markers being inversely proportional to their RMS. The \texttt{.par}-files of the new pulsars are taken from the published MPTA and PPTA data sets \citep{MPTA2024_data+noise,PPTA_GWB}, hence we will refer to these as the \textsc{IPTA20} data set. The \textsc{EPTA20} pulsars are placed at \SI{1}{\kilo\parsec} distance, the added \textsc{IPTA20} pulsars are located at a distance of $\sim\SI{4}{\kilo\parsec}$.  As any addition of pulsars to a data set increases the S/N of the CGW, we investigate two scenarios: (a) the added pulsars increase the S/N of the signal by $\sim$15\%, (b) the added pulsars increase the S/N by $\sim$60\%. The difference in the S/N was achieved by choosing the RMS of the added pulsars accordingly, a larger RMS means lower sensitivity. The cornerplot of the posterior distribution for the latter scenario (60\% S/N increase) is shown in Fig.~\ref{fig:epta-addon}. In both scenarios we find that the larger data set provides a better position recovery, and a more accurate strain recovery. On the other hand, the recovery of $f_\GW$ is still slightly impaired due to the significant contribution from the close-by pulsars. Comparing the two scenarios, we overall find that if the more distant pulsars contribute similar or slightly less S/N than the close-by pulsars, they still improve the parameter recovery.

\subsection{Bias in source position recovery}

\begin{figure}
    \centering
    \includegraphics[width=0.9\linewidth]{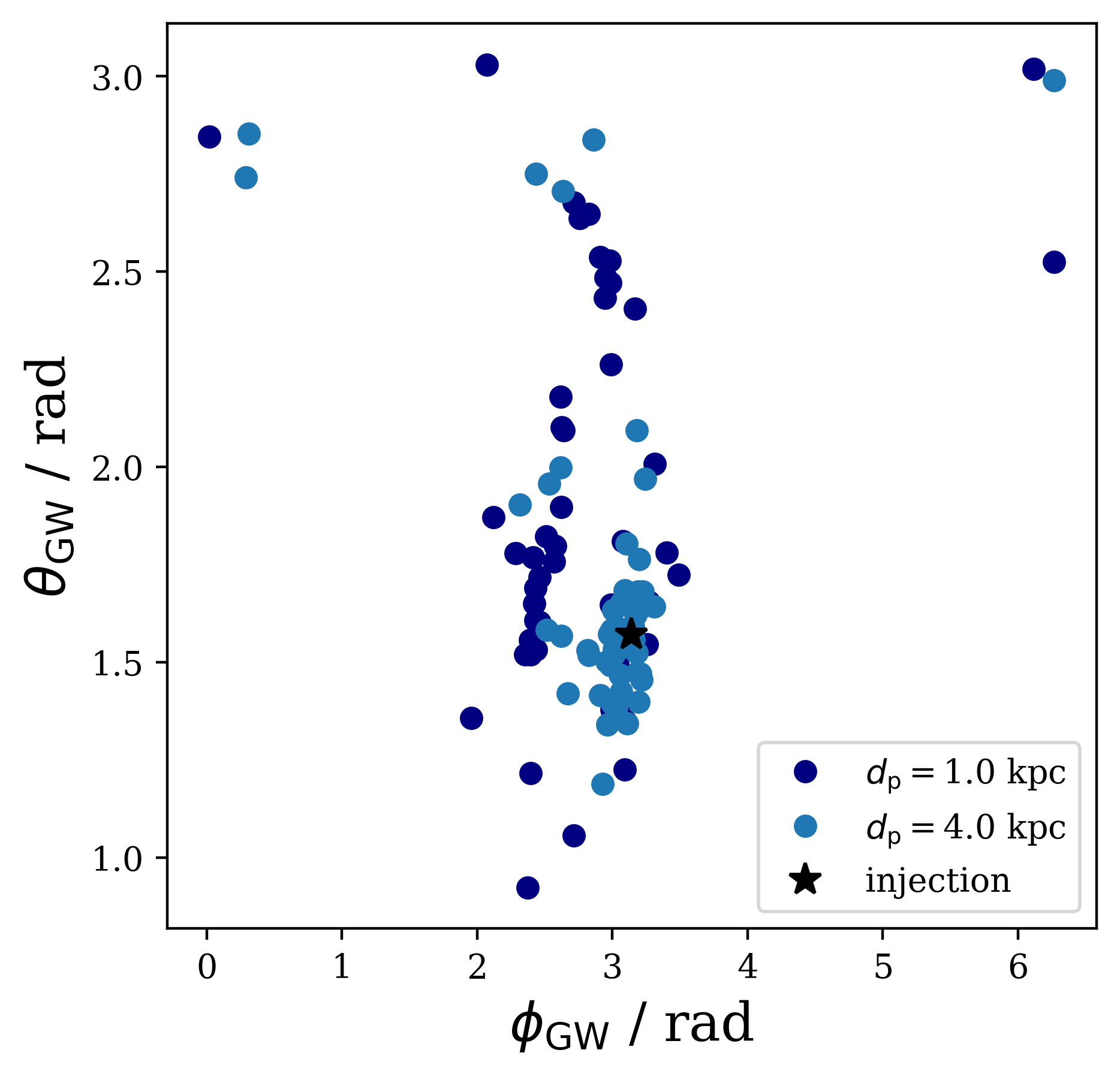}
    \caption{Maximum likelihood solutions for the CGW position parameters $\phi_\GW$ and $\theta_\GW$ (given in radians), using 50 realisations of a CGW signal injected to the \textsc{4PSR-galactic} data set with pulsars at a mean distance of $\Bar{d}_\mathrm{p} = \SI{1}{\kilo\parsec}$ and $\Bar{d}_\mathrm{p} = \SI{1}{\kilo\parsec}$. The injected source position is indicated with the black star.}
    \label{fig:4psr-gal-maxlike-CGWpos}
\end{figure}

As indicated by previous studies such as \cite{Zhu_2016}, we find that biases of parameter recoveries such as the CGW source position are closely linked to the confusion between the ET and PTs that is unaccounted for in an ET-only search. We demonstrate that an ET-only search with resolved pulsar terms is similarly capable of removing the position recovery bias as including the PTs in the search, using our miniature toy-model PTA. To this end, we inject a CGW with large S/N ($\lmc=9.0$, $d_L = \SI{15}{\mega\parsec}$) into four pulsars placed into the Galactic plane, with an RMS of \SI{100}{\nano\second}. These four pulsars sufficiently recreate the setup from \cite{Zhu_2016}, as their data set was also dominated by such a small number of pulsars with high ToA precision. The resulting maximum likelihood values for the CGW source position obtained from 50 exemplary realisations of each data set is shown in Fig.~\ref{fig:4psr-gal-maxlike-CGWpos}.

The source positions recovered using the realisations of the \SI{1}{\kilo\parsec} data set exhibit a similar offset to the injected source position as shown in Fig.~1 in \cite{Zhu_2016}. The slight differences between the distribution of the results in this work and in \cite{Zhu_2016} is caused by the different analysis methods: we present the maximum likelihood solutions obtained from a Bayesian sampling analysis, while \cite{Zhu_2016} uses a Frequentist method. 

Using the more distant data set, we can demonstrate that improving the source recovery is not only possible by including the PTs, as indicated by \cite{Zhu_2016}, but also by using good pulsars at a larger distance, such that the PTs become resolved.


\section{Variance of the Per-frequency Optimal Statistic}
\label{sec:PFOS}

A computationally efficient way to determine the presence of an HD correlation in a PTA data set is the frequentist Optimal Statistic (OS) \cite{Anholm_2009,Demorest_2013,Chamberlin_2015,Vigeland_2018}. Recently, \cite{Gersbach_2025} generalised this framework to the so-called Per-Frequency Optimal Statistic (PFOS), which allows us to characterise the spectral shape of a GW signal present in the PTA data set by determining the HD content in each frequency bin individually. 
\cite{Gersbach_2025} specifically showed that this method is also capable of indicating the presence of HD correlated signals limited to a few or a single frequency bins, such as a CGW. These narrow-band signals manifest in terms of significantly larger PSD estimators at these specific frequencies compared to the PSD estimators at other frequencies.

While the OS analysis framework per se is not the optimal way to detect a deterministic signal such as a CGW -- this would rather be a matched filter such as the $\mathcal{F}_\mathrm{e}$ or $\mathcal{F}_\mathrm{p}$ statistic \citep{BabakSesana2012} -- the ETs of the CGW signal exhibit an HD correlation \citep{RomanoAllen_2024}, which can show in an OS spectrum calculated using the PFOS. The OS analysis is typically employed as a first step in a PTA data analysis, as it is much faster and less computationally costly than a full Bayesian analysis of the data set, especially in light of ever growing data sets. Thus, it is interesting to explore the behaviour of this statistic in the presence of a CGW signal and varying pulsar distances.

\subsection{Analysis procedure}
\begin{figure}
    \centering
    \includegraphics[width=\linewidth]{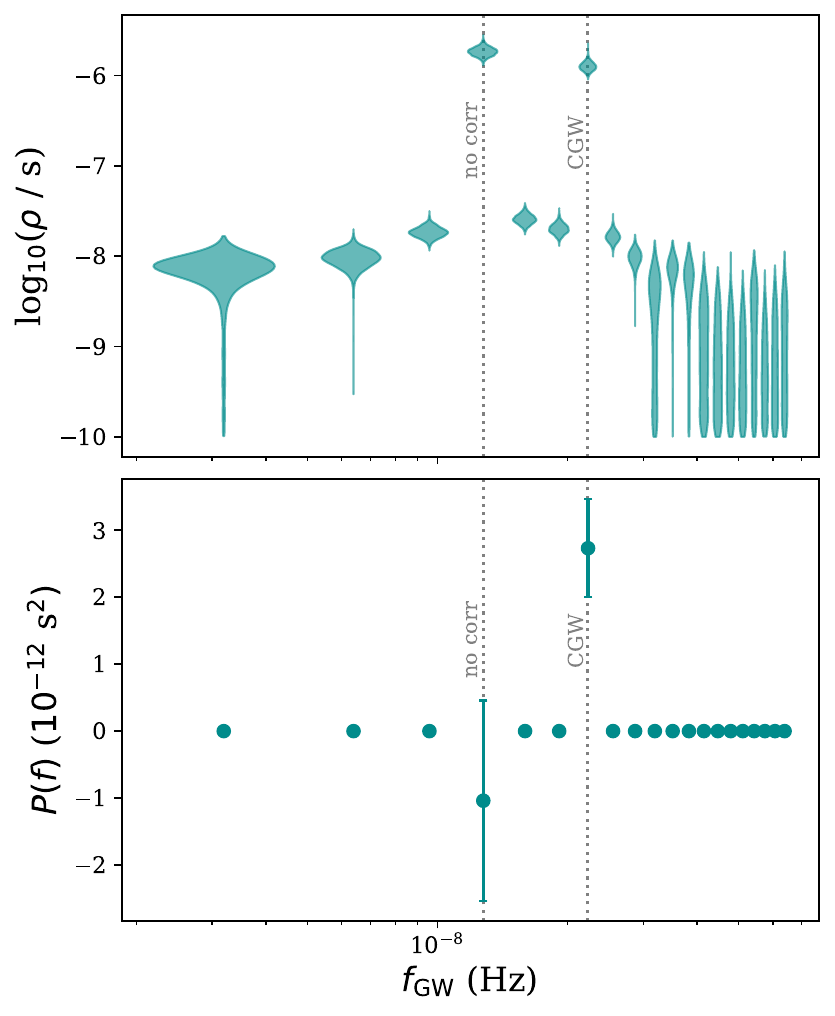}
    \caption{Comparison between a free spectrum and optimal statistic analysis at hand of a simulated data set with an injection of a CGW and a single bin uncorrelated CRN. Upper plot: Free spectrum, lower plot: per-frequency OS. }
    \label{fig:CGW_RN_comparison}
\end{figure}

The calculation of the PFOS requires the calculation of the PTA free spectrum, i.e.\ determining the amplitudes of the individual Fourier components of the cross-correlation matrix, Eq.~\eqref{eq:cc-matrix}.
The signal of a detectable circular binary manifests itself as a stand-out peak in a single bin of this free spectrum.
But the free spectrum does not yield any information about the spatial correlation of the signal. In Fig.~\ref{fig:CGW_RN_comparison} we present the free spectrum of a data set containing a single CGW signal at \SI{22.3}{\nano\hertz} and an uncorrelated single-bin red noise signal at \SI{11.2}{\nano\hertz} (both frequencies are multiples of $1/T\mathrm{obs}$). Both create a peak in the free spectrum, but only one originates from a CGW. 

Under the assumption that the data set only contains a GW signal in a few of its frequency bins, it can be analysed with the narrow-band PFOS \citep{Gersbach_2025}. Neglecting the pulsar pair covariance, i.e.\ the off-diagonal entries in $C_{ab,cd}$ in \cite{Gersbach_2025}, the estimator for the GW PSD at the $k^\mathrm{th}$ frequency bin with GW frequency $f_k$ is given as
\begin{equation}
    S(f_k) = \frac{\sum_{a<b}\Gamma_{ab}\sigma_{ab,k}^{-2}\rho_{ab,k}}{\sum_{a<b}\Gamma_{ab}^2\sigma_{ab}^{-2}},
\end{equation}
where, $\rho_{ab}(f_k)$ and $\sigma_{ab}(f_k)$ and denote the pulsar pair correlations and their uncertainties as given by Eq.~(33) and (34) in \cite{Gersbach_2025}.

As shown in the lower plot of Fig.~\ref{fig:CGW_RN_comparison}, this PFOS evaluation now allows us to identify the CGW as a narrow-band HD correlated signal, while the uncorrelated signal is not visible in the PFOS spectrum.

Across all evaluated frequency bins, the measured PFOS values depend on the specific realisation of the noise processes making up the analysed data set. Thus, when creating a series of data set realisations using the same noise parameters, the resulting PFOS values will vary, and the spread across all realisations indicates the reliability of the estimator.

As shown by \cite{Allen_2023}, the PT leads to an increase in the variance of the HD correlation. In an PFOS analysis, we therefore expect it to act like the uncorrelated narrow-band noise in Fig.~\ref{fig:CGW_RN_comparison}, increasing the uncertainty of the estimator and subsequently leading to a larger spread in the estimators across multiple data set realisations in the bins closest to the PT frequencies. In the data set of nearby pulsars, the ET and PT frequencies fall into the same frequency bin, meaning that the PT blurs the information content of the ET. In a data set with resolved ET and PTs, we expect on the other hand, that the separation clears the ET frequency bin from the PT noise, which in return should lead to a measurable decrease in the variation of the PFOS value in the ET bin.

\subsection{Results from the EPTA-like data set} 

\begin{figure}
    \centering\includegraphics[width=\linewidth]{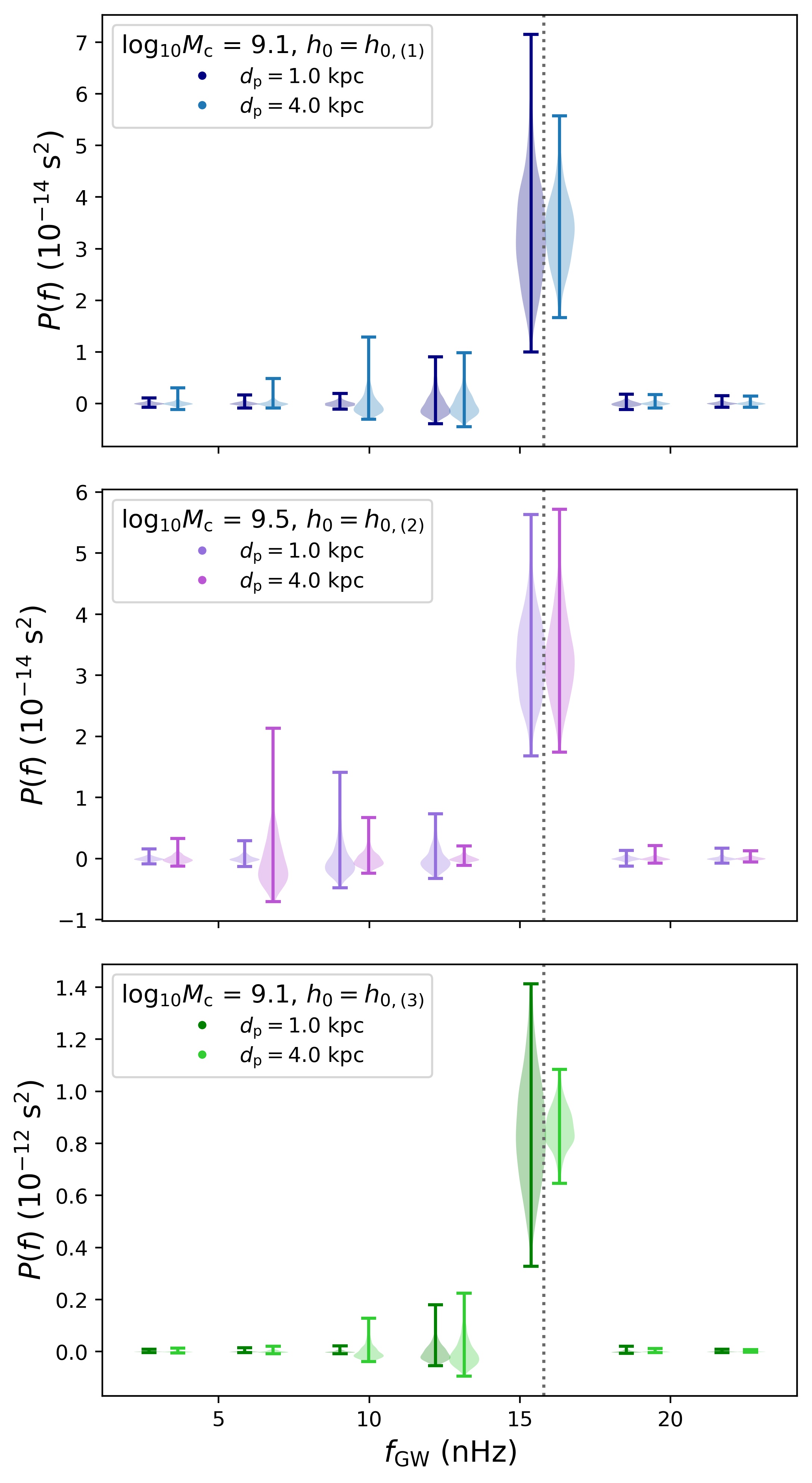}
    \caption{Distribution of narrow-band PFOS values for 500 realisations of the \textsc{EPTA20} PTA data set at different average pulsar distances.  Each point shows the PFOS estimator of the power spectrum amplitude at the respective frequency for a single data set realisation.  From top to bottom: Cases (1) - (3) as described in Sec.~\ref{ssec:simulation}}
    \label{fig:isotropic-PFOS_realisations}
\end{figure}

The results of the PFOS analyses for the 500 realisations of the \textsc{EPTA20} data sets with each of the three CGW signals described in Sec.~\ref{ssec:simulation} are shown in Fig.~\ref{fig:isotropic-PFOS_realisations}. The violins resemble the distribution of $P(f)$ estimates from each realisation. The comparisons demonstrate that the variation in the ET bin diminishes notably when moving to a data set that consists of pulsars further away.

For both simulations with $\lmc=9.1$, we clearly find that the spread between the realisations decreases as the ET and PT become resolved. This effect is more pronounced for a higher CGW strain (Case 3). Simultaneously we notice that the spread across the realisations increases in the adjacent lower frequency bin. This is testament to the `noise-like' treatment of the PT. As the ET falls together with the PT, the latter one adds noise to the signal present in the corresponding frequency bin. But with increasing separation between both, the noise shifts towards the lower frequency bin, leaving only the signal in the ET frequency bin. 
Unsurprisingly, the spread in the OS values of the simulations using $\lmc=9.5$ (Case 2) does not differ significantly for the two mean pulsar distances. As discussed previously, for $\lmc=9.5$, the ET and PT are resolved in all pulsars already for $\Bar{d}_\mathrm{p}=\SI{1}{\kilo\parsec}$, and the further separation of the PTs from the ET does not improve the PFOS S/N in the ET frequency bin.

Overall we deduce from the results of both PTA setups, that the identification of narrowband HD correlations from CGW signals using the PFOS method benefits from separating the ET and PT frequencies by moving to more distant pulsars. This can be of special interest for lower GW strains, i.e.\ sources at the detection limit: the CGW would be more reliably visible as a raised $P(f)$ value in the PFOS spectrum for pulsars at a large distance to the Earth, while a larger spread for closer pulsars would mean that the $P(f)$ is more likely to vanish in the noise floor of the data.

\section{Conclusions and Outlook}
\label{sec:conclusion}

We have shown, that including pulsars at a larger distance to the Earth can improve the performance of two PTA CGW analysis tools in the presence of a high frequency ($\gtrsim \SI{10}{\nano\hertz}$) CGW signal within the respective data set, due to the separation of the ET and the PT.

The Bayesian likelihood search using an ET-only CGW model improves using pulsars at a greater distance. Especially, the posterior distributions of the source position parameters, the GW frequency and the strain become more constrained. We tested this effect using an idealized scenario with pulsars either nearby or further away, but we have also demonstrated that adding more distant pulsars to an existing array improves the capabilities of the PTA to constrain CGW parameters in an ET-only search.

Our analyses demonstrated that the improvement of this technique is highly dependent on the constellation of pulsars with respect to the CGW source position, as well as the S/N with which the signal appears in each pulsar's residuals. While we did not check this implicitly for the frequentist analysis, they behave similarly, so it is safe to assume that it applies to both techniques.

The frequentist analysis with the narrowband optimal statistic \citep{Gersbach_2025} more reliably indicates the excess power in the regarding frequency bins close to the GW frequency if applied to pulsars with resolved PTs. This effect is especially noteworthy for weaker GW sources, where the CGW is close to the noise floor of the timing residuals. In the (currently unrealistic) case that the CGW strain is significantly above the noise floor, this separation effect proves not as relevant for the performance of the PFOS, as the CGW is always visible.

Overall, we have explored an alternative approach to tackle the impairment that CGW search and analysis techniques suffer from, in light of unconstrained PTs due to imprecisely known pulsar distances. We have demonstrated, that parameter misspecifications and loss of S/N can be diminished by using pulsars with resolved PTs.

The findings of this work also have several subsequent implications. The notable improvement that a group of pulsars with fully resolved PTs can provide, suggests that it may be interesting for the analysis of a future PTA data set with $\mathcal{O}(100+)$ pulsars, to perform a CGW search only on a subset of pulsars at a larger distance. This could involve improving the detection of a high-frequency source using the PFOS, or improving the constraints on the parameters of a CGW once one is detected. Additionally, the location of the example SMBHB candidates in Fig.~\ref{fig:flmc_parameterspace} indicates, that there are some high-frequency candidates, for which an ET-only targeted search is likely sufficient, whereas others with a predicted $f_\GW$ of only a few nano-Hertz should be targeted with a full ET+PT model.
Furthermore, the behaviour of the GW frequency posterior (cf.\ Fig.~\ref{fig:epta20-case2and3}), namely that it sometimes breaks out to the PT frequencies, might be a usable feature to constrain the frequency evolution of the CGW without the need of precisely known pulsar distances.

\begin{acknowledgements}
We thank Mikel Falxa and Stanislav Babak for the fruitful discussions and comments on CGW modelling and the Optimal Statistic.

All authors affiliated with the Max-Planck-Gesellschaft (MPG) acknowledge its constant support.
KG acknowledges support from the International Max Planck Research School (IMPRS) for Astronomy and Astrophysics at the Universities of Bonn and Cologne and the Bonn-Cologne Graduate School of Physics and Astronomy.

The analysis done in this publication made use of the open source pulsar analysis packages \textsc{tempo2} \citep{tempo2}, \textsc{libstempo} \citep{libstempo}, \textsc{enterprise} \citep{enterprise} and \textsc{enterprise\_extensions} \citep{enterprise_extensions}, as well as open source \textsc{Python} libraries including Numpy \citep{numpy}, Matplotlib \citep{matplotlib}, Astropy \citep{astropy:2013,astropy:2018,astropy:2022} and Chainconsumer \citep{chainconsumer}.
\end{acknowledgements}

\bibliographystyle{aa}
\bibliography{bibliography}

@String{mnras = "Mon. Not. R. Ast. Soc."}

@String{aap = "Astron. Astrophys."}

@String{apj = "Astrophys. J."}

@String{apjl = "Astrophys. J. Lett."}

@String{prd = "Phys. Rev. D"}

@String{prl = "Phys. Rev. Lett."}

@String{nima = "Nucl. Instrum. Meth. A"}

@String{pasa = "Pub. Astron. Soc. Aust."}

@book{handbook,
	author= {D. R. Lorimer and M. Kramer},
	title = {Handbook of Pulsar Astronomy},
	publisher = {Cambridge University Press},
	year = {2005}
}

@article{Miles_2022,
    title={The MeerKAT Pulsar Timing Array: first data release},
    volume={519},
    number={3},
    journal=mnras,
    author={Miles, M T and Shannon, R M and Bailes, M and Reardon, D J and Keith, M J and Cameron, A D and Parthasarathy, A and Shamohammadi, M and Spiewak, R and van Straten, W and Buchner, S and Camilo, F and Geyer, M and Karastergiou, A and Kramer, M and Serylak, M and Theureau, G and Venkatraman Krishnan, V},
    year={2022},
    month=dec,
    pages={3976–3991},
    doi = {10.1093/mnras/stac3644}
}

@ARTICLE{MPTA2024_data+noise,
    author = {M Miles and others},
    title = {The MeerKAT Pulsar Timing Array: The 4.5-year data release and the noise and stochastic signals of the millisecond pulsar population},
    journal=mnras,
    year = 2024
}

@ARTICLE{EPTA_DR2_DATA,
       author = {{EPTA Collaboration} and {Antoniadis}, J. and {Babak}, S. and {Bak Nielsen}, A. -S. and {Bassa}, C.~G. and {Berthereau}, A. and {Bonetti}, M. and {Bortolas}, E. and {Brook}, P.~R. and {Burgay}, M. and {Caballero}, R.~N. and {Chalumeau}, A. and {Champion}, D.~J. and {Chanlaridis}, S. and {Chen}, S. and {Cognard}, I. and {Desvignes}, G. and {Falxa}, M. and {Ferdman}, R.~D. and {Franchini}, A. and {Gair}, J.~R. and {Goncharov}, B. and {Graikou}, E. and {Grie{\ss}meier}, J. -M. and {Guillemot}, L. and {Guo}, Y.~J. and {Hu}, H. and {Iraci}, F. and {Izquierdo-Villalba}, D. and {Jang}, J. and {Jawor}, J. and {Janssen}, G.~H. and {Jessner}, A. and {Karuppusamy}, R. and {Keane}, E.~F. and {Keith}, M.~J. and {Kramer}, M. and {Krishnakumar}, M.~A. and {Lackeos}, K. and {Lee}, K.~J. and {Liu}, K. and {Liu}, Y. and {Lyne}, A.~G. and {McKee}, J.~W. and {Main}, R.~A. and {Mickaliger}, M.~B. and {Ni{\c{t}}u}, I.~C. and {Parthasarathy}, A. and {Perera}, B.~B.~P. and {Perrodin}, D. and {Petiteau}, A. and {Porayko}, N.~K. and {Possenti}, A. and {Quelquejay Leclere}, H. and {Samajdar}, A. and {Sanidas}, S.~A. and {Sesana}, A. and {Shaifullah}, G. and {Speri}, L. and {Spiewak}, R. and {Stappers}, B.~W. and {Susarla}, S.~C. and {Theureau}, G. and {Tiburzi}, C. and {van der Wateren}, E. and {Vecchio}, A. and {Venkatraman Krishnan}, V. and {Verbiest}, J.~P.~W. and {Wang}, J. and {Wang}, L. and {Wu}, Z.},
        title = "{The second data release from the European Pulsar Timing Array. I. The dataset and timing analysis}",
      journal = {\aap},
         year = 2023,
        month = oct,
       volume = {678},
          eid = {A48},
        pages = {A48},
          doi = {10.1051/0004-6361/202346841}
}

@article{EPTA_DR2_GWB,
	author = {{EPTA Collaboration and InPTA Collaboration:} and {Antoniadis, J.} and {Arumugam, P.} and {Arumugam, S.} and {Babak, S.} and {Bagchi, M.} and {Bak Nielsen, A.-S.} and {Bassa, C. G.} and {Bathula, A.} and {Berthereau, A.}},
	title = {The second data release from the European Pulsar Timing Array - III. Search for gravitational wave signals},
	journal = aap,
	year = 2023,
	volume = 678,
	pages = "A50",
    doi = {10.1051/0004-6361/202346844}
}

@ARTICLE{NG15_GWB,
    author = {{Agazie}, Gabriella and {Anumarlapudi}, Akash and {Archibald}, Anne M. and {Arzoumanian}, Zaven and {Baker}, Paul T. and {B{\'e}csy}, Bence and {Blecha}, Laura and {Brazier}, Adam and {Nanograv Collaboration}},
    title = "{The NANOGrav 15 yr Data Set: Evidence for a Gravitational-wave Background}",
    journal = apjl,
    year = 2023,
    month = jul,
    volume = {951},
    number = {1},
    pages = {L8},
    doi = {10.3847/2041-8213/acdac6}
}

@ARTICLE{PPTA_GWB,
    author = {{Reardon}, Daniel J. and {Zic}, Andrew and {Shannon}, Ryan M. and {Hobbs}, George B. and {Bailes}, Matthew and {Di Marco}, Valentina and {Kapur}, Agastya and {Rogers}, Axl F. and {Thrane}, Eric and {Askew}, Jacob and {Bhat}, N.~D. Ramesh and {Cameron}, Andrew and {Cury{\l}o}, Ma{\l}gorzata and {Coles}, William A. and {Dai}, Shi and {Goncharov}, Boris and {Kerr}, Matthew and {Kulkarni}, Atharva and {Levin}, Yuri and {Lower}, Marcus E. and {Manchester}, Richard N. and {Mandow}, Rami and {Miles}, Matthew T. and {Nathan}, Rowina S. and {Os{\l}owski}, Stefan and {Russell}, Christopher J. and {Spiewak}, Ren{\'e}e and {Zhang}, Songbo and {Zhu}, Xing-Jiang},
    title = "{Search for an Isotropic Gravitational-wave Background with the Parkes Pulsar Timing Array}",
    journal = apjl,
    year = 2023,
    month = jul,
    volume = {951},
    number = {1},
    pages = {L6},
    doi = {10.3847/2041-8213/acdd02}
}

@ARTICLE{CPTA_GWB,
    author = {{Xu}, Heng and {Chen}, Siyuan and {Guo}, Yanjun and {Jiang}, Jinchen and {Wang}, Bojun and {Xu}, Jiangwei and {Xue}, Zihan and {Nicolas Caballero}, R. and {Yuan}, Jianping and {Xu}, Yonghua and {Wang}, Jingbo and {Hao}, Longfei and {Luo}, Jingtao and {Lee}, Kejia and {Han}, Jinlin and {Jiang}, Peng and {Shen}, Zhiqiang and {Wang}, Min and {Wang}, Na and {Xu}, Renxin and {Wu}, Xiangping and {Manchester}, Richard and {Qian}, Lei and {Guan}, Xin and {Huang}, Menglin and {Sun}, Chun and {Zhu}, Yan},
    title = "{Searching for the Nano-Hertz Stochastic Gravitational Wave Background with the Chinese Pulsar Timing Array Data Release I}",
    journal = {Research in Astronomy and Astrophysics},
    year = 2023,
    month = jul,
    volume = {23},
    number = {7},     
    pages = {075024},
    doi = {10.1088/1674-4527/acdfa5}
}

@ARTICLE{IPTA_2013,
       author = {{Manchester}, R.~N. and {IPTA}},
        title = "{The International Pulsar Timing Array}",
      journal = {Classical and Quantum Gravity},
         year = 2013,
        month = nov,
       volume = {30},
       number = {22},
          eid = {224010},
        pages = {224010},
          doi = {10.1088/0264-9381/30/22/224010},
archivePrefix = {arXiv},
       eprint = {1309.7392}
}

@ARTICLE{IPTA_DR2_CGW,
       author = {{Falxa}, M. and {Babak}, S. and {Baker}, P.~T. and {B{\'e}csy}, B. and {Chalumeau}, A. and {Chen}, S. and {Chen}, Z. and {Cornish}, N.~J. and {Guillemot}, L. and {Hazboun}, J.~S. and {Mingarelli}, C.~M.~F. and {Parthasarathy}, A. and {IPTA Collaboration}},
        title = "{Searching for continuous Gravitational Waves in the second data release of the International Pulsar Timing Array}",
      journal = {\mnras},
         year = 2023,
        month = jun,
       volume = {521},
       number = {4},
        pages = {5077-5086},
          doi = {10.1093/mnras/stad812},
archivePrefix = {arXiv},
       eprint = {2303.10767},
 primaryClass = {gr-qc},
       adsurl = {https://ui.adsabs.harvard.edu/abs/2023MNRAS.521.5077F}
}

@ARTICLE{Anholm_2009,
       author = {{Anholm}, Melissa and {Ballmer}, Stefan and {Creighton}, Jolien D.~E. and {Price}, Larry R. and {Siemens}, Xavier},
        title = "{Optimal strategies for gravitational wave stochastic background searches in pulsar timing data}",
      journal = {\prd},
         year = 2009,
        month = apr,
       volume = {79},
       number = {8},
          eid = {084030},
        pages = {084030},
          doi = {10.1103/PhysRevD.79.084030}
}

@ARTICLE{AllenRomano_1999,
       author = {{Allen}, Bruce and {Romano}, Joseph D.},
        title = "{Detecting a stochastic background of gravitational radiation: Signal processing strategies and sensitivities}",
      journal = {\prd},
         year = 1999,
        month = may,
       volume = {59},
       number = {10},
        pages = {102001},
          doi = {10.1103/PhysRevD.59.102001},
       eprint = {gr-qc/9710117},
 primaryClass = {gr-qc}
}

@ARTICLE{Allen_2023,
       author = {{Allen}, Bruce},
        title = "{Variance of the Hellings-Downs correlation}",
      journal = {\prd},
         year = 2023,
        month = feb,
       volume = {107},
       number = {4},
        pages = {043018},
          doi = {10.1103/PhysRevD.107.043018}
}

@ARTICLE{Arzoumanian_2014,
       author = {{Arzoumanian}, Z. and {Brazier}, A. and {Burke-Spolaor}, S. and {Chamberlin}, S.~J. and {Chatterjee}, S. and {Cordes}, J.~M. and {Demorest}, P.~B. and {Deng}, X. and {Dolch}, T. and {Ellis}, J.~A. and {Ferdman}, R.~D. and {Garver-Daniels}, N. and {Jenet}, F. and {Jones}, G. and {Kaspi}, V.~M. and {Koop}, M. and {Lam}, M.~T. and {Lazio}, T.~J.~W. and {Lommen}, A.~N. and {Lorimer}, D.~R. and {Luo}, J. and {Lynch}, R.~S. and {Madison}, D.~R. and {McLaughlin}, M.~A. and {McWilliams}, S.~T. and {Nice}, D.~J. and {Palliyaguru}, N. and {Pennucci}, T.~T. and {Ransom}, S.~M. and {Sesana}, A. and {Siemens}, X. and {Stairs}, I.~H. and {Stinebring}, D.~R. and {Stovall}, K. and {Swiggum}, J. and {Vallisneri}, M. and {van Haasteren}, R. and {Wang}, Y. and {Zhu}, W.~W. and {NANOGrav Collaboration}},
        title = "{Gravitational Waves from Individual Supermassive Black Hole Binaries in Circular Orbits: Limits from the North American Nanohertz Observatory for Gravitational Waves}",
      journal = {\apj},
         year = 2014,
        month = oct,
       volume = {794},
       number = {2},
          eid = {141},
        pages = {141},
          doi = {10.1088/0004-637X/794/2/141},
archivePrefix = {arXiv},
       eprint = {1404.1267},
 primaryClass = {astro-ph.GA},
       adsurl = {https://ui.adsabs.harvard.edu/abs/2014ApJ...794..141A}
}

@ARTICLE{Babak_2016,
       author = {{Babak}, S. and {Petiteau}, A. and {Sesana}, A. and {Brem}, P. and {Rosado}, P.~A. and {Taylor}, S.~R. and {Lassus}, A. and {Hessels}, J.~W.~T. and {Bassa}, C.~G. and {Burgay}, M. and {Caballero}, R.~N. and {Champion}, D.~J. and {Cognard}, I. and {Desvignes}, G. and {Gair}, J.~R. and {Guillemot}, L. and {Janssen}, G.~H. and {Karuppusamy}, R. and {Kramer}, M. and {Lazarus}, P. and {Lee}, K.~J. and {Lentati}, L. and {Liu}, K. and {Mingarelli}, C.~M.~F. and {Os{\l}owski}, S. and {Perrodin}, D. and {Possenti}, A. and {Purver}, M.~B. and {Sanidas}, S. and {Smits}, R. and {Stappers}, B. and {Theureau}, G. and {Tiburzi}, C. and {van Haasteren}, R. and {Vecchio}, A. and {Verbiest}, J.~P.~W.},
        title = "{European Pulsar Timing Array limits on continuous gravitational waves from individual supermassive black hole binaries}",
      journal = {\mnras},
         year = 2016,
        month = jan,
       volume = {455},
       number = {2},
        pages = {1665-1679},
          doi = {10.1093/mnras/stv2092},
       eprint = {1509.02165}
}

@ARTICLE{Babak_2024,
       author = {{Babak}, Stanislav and {Falxa}, Mikel and {Franciolini}, Gabriele and {Pieroni}, Mauro},
        title = "{Forecasting the sensitivity of pulsar timing arrays to gravitational wave backgrounds}",
      journal = {\prd},
         year = 2024,
        month = sep,
       volume = {110},
       number = {6},
          doi = {10.1103/PhysRevD.110.063022},
archivePrefix = {arXiv},
       eprint = {2404.02864},
       adsurl = {https://ui.adsabs.harvard.edu/abs/2024PhRvD.110f3022B}
}

@ARTICLE{BabakSesana2012,
       author = {{Babak}, Stanislav and {Sesana}, Alberto},
        title = "{Resolving multiple supermassive black hole binaries with pulsar timing arrays}",
      journal = {\prd},
         year = 2012,
        month = feb,
       volume = {85},
       number = {4},
          eid = {044034},
        pages = {044034},
          doi = {10.1103/PhysRevD.85.044034},
archivePrefix = {arXiv},
       eprint = {1112.1075},
 primaryClass = {astro-ph.CO},
       adsurl = {https://ui.adsabs.harvard.edu/abs/2012PhRvD..85d4034B}
}

@ARTICLE{Bailes_Rev2021,
       author = {{Bailes}, M. and {Berger}, B.~K. and {Brady}, P.~R. and {Branchesi}, M. and {Danzmann}, K. and {Evans}, M. and {Holley-Bockelmann}, K. and {Iyer}, B.~R. and {Kajita}, T. and {Katsanevas}, S. and {Kramer}, M. and {Lazzarini}, A. and {Lehner}, L. and {Losurdo}, G. and {L{\"u}ck}, H. and {McClelland}, D.~E. and {McLaughlin}, M.~A. and {Punturo}, M. and {Ransom}, S. and {Raychaudhury}, S. and {Reitze}, D.~H. and {Ricci}, F. and {Rowan}, S. and {Saito}, Y. and {Sanders}, G.~H. and {Sathyaprakash}, B.~S. and {Schutz}, B.~F. and {Sesana}, A. and {Shinkai}, H. and {Siemens}, X. and {Shoemaker}, D.~H. and {Thorpe}, J. and {van den Brand}, J.~F.~J. and {Vitale}, S.},
        title = "{Gravitational-wave physics and astronomy in the 2020s and 2030s}",
      journal = {Nature Reviews Physics},
         year = 2021,
        month = jan,
       volume = {3},
       number = {5},
        pages = {344-366},
          doi = {10.1038/s42254-021-00303-8},
       adsurl = {https://ui.adsabs.harvard.edu/abs/2021NatRP...3..344B}
}

@ARTICLE{Begelman_1980,
    author = {{Begelman}, M.~C. and {Blandford}, R.~D. and {Rees}, M.~J.},
    title = "{Massive black hole binaries in active galactic nuclei}",
    journal = {\nat},
    year = 1980,
    month = sep,
    volume = {287},
    number = {5780},
    pages = {307-309},
    doi = {10.1038/287307a0}
}

@ARTICLE{BurkeSpolaor_2015,
       author = {{Burke-Spolaor}, Sarah},
        title = "{Gravitational-Wave Detection and Astrophysics with Pulsar Timing Arrays}",
      journal = {arXiv e-prints},
         year = 2015,
        month = nov,
          eid = {arXiv:1511.07869},
        pages = {arXiv:1511.07869},
          doi = {10.48550/arXiv.1511.07869},
archivePrefix = {arXiv},
       eprint = {1511.07869},
 primaryClass = {astro-ph.IM},
       adsurl = {https://ui.adsabs.harvard.edu/abs/2015arXiv151107869B}
}

@ARTICLE{CardinalTremblay_2025,
       author = {{Cardinal Tremblay}, Jacob and {Goncharov}, Boris and {van Haasteren}, Rutger and {Bhat}, N.~D. Ramesh and {Chen}, Zu-Cheng and {Di Marco}, Valentina and {Iguchi}, Satoru and {Kapur}, Agastya and {Ling}, Wenhua and {Mandow}, Rami and {Mishra}, Saurav and {Reardon}, Daniel J. and {Shannon}, Ryan M. and {Sudou}, Hiroshi and {Wang}, Jingbo and {Zhao}, Shi-Yi and {Zhu}, Xing-Jiang and {Zic}, Andrew},
        title = "{A Multi-Messenger Search for the Supermassive Black Hole Binary in 3C 66B with the Parkes Pulsar Timing Array}",
      journal = {arXiv e-prints},
         year = 2025,
        month = aug,
        pages = {arXiv:2508.20007},
          doi = {10.48550/arXiv.2508.20007}
}

@ARTICLE{Chamberlin_2015,
    author = {{Chamberlin}, Sydney J. and {Creighton}, Jolien D.~E. and {Siemens}, Xavier and {Demorest}, Paul and {Ellis}, Justin and {Price}, Larry R. and {Romano}, Joseph D.},
    title = "{Time-domain implementation of the optimal cross-correlation statistic for stochastic gravitational-wave background searches in pulsar timing data}",
    journal = prd,
    year = 2015,
    month = feb,
    volume = {91},
    number = {4},
    pages = {044048},
    doi = {10.1103/PhysRevD.91.044048}
}

@ARTICLE{Chen_2022,
       author = {{Chen}, Jie-Wen and {Wang}, Yan},
        title = "{Parameter-estimation Biases for Eccentric Supermassive Binary Black Holes in Pulsar Timing Arrays: Biases Caused by Ignored Pulsar Terms}",
      journal = {\apj},
         year = 2022,
        month = apr,
       volume = {929},
       number = {2},
          eid = {168},
        pages = {168},
          doi = {10.3847/1538-4357/ac5bd4}
}

@ARTICLE{Corbin2010,
       author = {{Corbin}, Vincent and {Cornish}, Neil J.},
        title = "{Pulsar Timing Array Observations of Massive Black Hole Binaries}",
      journal = {arXiv e-prints},
         year = 2010,
        month = aug,
          eid = {arXiv:1008.1782},
        pages = {arXiv:1008.1782},
          doi = {10.48550/arXiv.1008.1782},
       eprint = {1008.1782}
}

@INPROCEEDINGS{Deller_2011,
       author = {{Deller}, A.~T. and {Brisken}, W.~F. and {Chatterjee}, S. and {Cordes}, J.~M. and {Goss}, W.~M. and {Janssen}, G.~H. and {Kovalev}, Y.~Y. and {Lazio}, T.~J.~W. and {Petrov}, L. and {Stappers}, B.~W.},
        title = "{PSR{\ensuremath{\pi}}: A large VLBA pulsar astrometry program}",
    booktitle = {20th Meeting of the European VLBI Group for Geodesy and Astronomy},
         year = 2011,
       editor = {{Alef}, Walter and {Bernhart}, Simone and {Nothnagel}, Axel},
        month = jul,
        pages = {178-182},
          doi = {10.48550/arXiv.1110.1979},
archivePrefix = {arXiv},
       eprint = {1110.1979},
 primaryClass = {astro-ph.SR},
       adsurl = {https://ui.adsabs.harvard.edu/abs/2011evga.conf..178D}
}

@ARTICLE{Demorest_2013,
    author = {{Demorest}, P.~B. and {Ferdman}, R.~D. and {Gonzalez}, M.~E. and {Nice}, D. and {Ransom}, S. and {Stairs}, I.~H. and {Arzoumanian}, Z. and {Brazier}, A. and {Burke-Spolaor}, S. and {Chamberlin}, S.~J. and {Cordes}, J.~M. and {Ellis}, J. and {Finn}, L.~S. and {Freire}, P. and {Giampanis}, S. and {Jenet}, F. and {Kaspi}, V.~M. and {Lazio}, J. and {Lommen}, A.~N. and {McLaughlin}, M. and {Palliyaguru}, N. and {Perrodin}, D. and {Shannon}, R.~M. and {Siemens}, X. and {Stinebring}, D. and {Swiggum}, J. and {Zhu}, W.~W.},
    title = "{Limits on the Stochastic Gravitational Wave Background from the North American Nanohertz Observatory for Gravitational Waves}",
    journal = apj,
    year = 2013,
    month = jan,
    volume = {762},
    number = {2},
    pages = {94},
    doi = {10.1088/0004-637X/762/2/94}
}

@ARTICLE{Detweiler_1979,
       author = {{Detweiler}, S.},
        title = "{Pulsar timing measurements and the search for gravitational waves}",
      journal = {\apj},
         year = 1979,
        month = dec,
       volume = {234},
        pages = {1100-1104},
          doi = {10.1086/157593},
       adsurl = {https://ui.adsabs.harvard.edu/abs/1979ApJ...234.1100D}
}

@ARTICLE{Ellis_2012,
       author = {{Ellis}, J.~A. and {Jenet}, F.~A. and {McLaughlin}, M.~A.},
        title = "{Practical Methods for Continuous Gravitational Wave Detection Using Pulsar Timing Data}",
      journal = {\apj},
         year = 2012,
        month = jul,
       volume = {753},
       number = {2},
          eid = {96},
        pages = {96},
          doi = {10.1088/0004-637X/753/2/96},
archivePrefix = {arXiv},
       eprint = {1202.0808}
}

@ARTICLE{EstabrookWahlquist_1978,
       author = {{Estabrook}, F.~B. and {Wahlquist}, H.~D.},
        title = "{Prospects for detection of gravitational radiation by simultaneous Doppler tracking of several spacecraft.}",
      journal = {Acta Astronautica},
         year = 1978,
        month = feb,
       volume = {5},
        pages = {5-7},
          doi = {10.1016/0094-5765(78)90030-9},
       adsurl = {https://ui.adsabs.harvard.edu/abs/1978AcAau...5....5E}
}

@ARTICLE{Foster_1990,
    author = {{Foster}, R.~S. and {Backer}, D.~C.},
    title = "{Constructing a Pulsar Timing Array}",
    journal = {\apj},
    year = 1990,
    month = sep,
    volume = {361},
    pages = {300},
    doi = {10.1086/169195}
}

@ARTICLE{Gair_2014,
       author = {{Gair}, Jonathan and {Romano}, Joseph D. and {Taylor}, Stephen and {Mingarelli}, Chiara M.~F.},
        title = "{Mapping gravitational-wave backgrounds using methods from CMB analysis: Application to pulsar timing arrays}",
      journal = {\prd},
         year = 2014,
        month = oct,
       volume = {90},
       number = {8},
          eid = {082001},
          doi = {10.1103/PhysRevD.90.082001}
}

@ARTICLE{Gersbach_2025,
       author = {{Gersbach}, Kyle A. and {Taylor}, Stephen R. and {Meyers}, Patrick M. and {Romano}, Joseph D.},
        title = "{Spatial and spectral characterization of the gravitational-wave background with the PTA optimal statistic}",
      journal = {\prd},
         year = 2025,
        month = jan,
       volume = {111},
       number = {2},
          eid = {023027},
          doi = {10.1103/PhysRevD.111.023027}
}

@ARTICLE{HellingsDowns_1983,
    author = {{Hellings}, R.~W. and {Downs}, G.~S.},
    title = "{Upper limits on the isotropic gravitational radiation background from pulsar timing analysis.}",
    journal = {\apjl},
    year = 1983,
    month = feb,
    volume = {265},
    pages = {L39-L42},
    doi = {10.1086/183954}
}

@ARTICLE{Jaffe_2003,
       author = {{Jaffe}, A.~H. and {Backer}, D.~C.},
        title = "{Gravitational Waves Probe the Coalescence Rate of Massive Black Hole Binaries}",
      journal = {\apj},
         year = 2003,
        month = feb,
       volume = {583},
       number = {2},
        pages = {616-631},
          doi = {10.1086/345443}
}

@ARTICLE{Kato_2023,
       author = {{Kato}, Ryo and {Takahashi}, Keitaro},
        title = "{Precision of localization of single gravitational-wave source with pulsar timing array}",
      journal = {\prd},
         year = 2023,
        month = dec,
       volume = {108},
       number = {12},
          eid = {123535},
          doi = {10.1103/PhysRevD.108.123535}
}

@ARTICLE{Komossa_2023,
       author = {{Komossa}, S. and {Grupe}, D. and {Kraus}, A. and {Gurwell}, M.~A. and {Haiman}, Z. and {Liu}, F.~K. and {Tchekhovskoy}, A. and {Gallo}, L.~C. and {Berton}, M. and {Blandford}, R. and {G{\'o}mez}, J.~L. and {Gonzalez}, A.~G.},
        title = "{Absence of the predicted 2022 October outburst of OJ 287 and implications for binary SMBH scenarios}",
      journal = {\mnras},
         year = 2023,
        month = jun,
       volume = {522},
       number = {1},
        pages = {L84-L88},
          doi = {10.1093/mnrasl/slad016}
}

@ARTICLE{Kormendy_1995,
    author = {{Kormendy}, John and {Richstone}, Douglas},
    title = "{Inward Bound---The Search For Supermassive Black Holes In Galactic Nuclei}",
    journal = {\araa},
    year = 1995,
    month = jan,
    volume = {33},
    pages = {581},
    doi = {10.1146/annurev.aa.33.090195.003053}
}

@ARTICLE{Lee_2011,
       author = {{Lee}, K.~J. and {Wex}, N. and {Kramer}, M. and {Stappers}, B.~W. and {Bassa}, C.~G. and {Janssen}, G.~H. and {Karuppusamy}, R. and {Smits}, R.},
        title = "{Gravitational wave astronomy of single sources with a pulsar timing array}",
      journal = {\mnras},
         year = 2011,
        month = jul,
       volume = {414},
       number = {4},
        pages = {3251-3264},
          doi = {10.1111/j.1365-2966.2011.18622.x},
archivePrefix = {arXiv},
       eprint = {1103.0115},
 primaryClass = {astro-ph.HE},
       adsurl = {https://ui.adsabs.harvard.edu/abs/2011MNRAS.414.3251L}
}

@article{Maggiore_2000,
    author = "Michele Maggiore",
    title = "Gravitational Wave Experiments and Early Universe Cosmology",
    journal = "Phys. Rept.",
    volume = 331,
    pages = "283",
    year = 2000,
    doi = {10.1016/S0370-1573(99)00102-7}
}

@ARTICLE{Middleton2025,
       author = {{Middleton}, Hannah and {Shannon}, Ryan M. and {Bailes}, Matthew and {Cameron}, Andrew D. and {Corongiu}, Alessandro and {Geyer}, Marisa and {Jones}, Max and {Kramer}, Michael and {Miles}, Matthew T. and {Parthasarathy}, Aditya and {Possenti}, Andrea and {Reardon}, Daniel J.},
        title = "{A simple optimization for the MeerKAT pulsar timing array}",
      journal = {\mnras},
         year = 2025,
        month = jun,
       volume = {540},
       number = {1},
        pages = {603-611},
          doi = {10.1093/mnras/staf748},
       adsurl = {https://ui.adsabs.harvard.edu/abs/2025MNRAS.540..603M}
}

@ARTICLE{Milosavljevic_2001,
       author = {{Milosavljevi{\'c}}, Milo{\v{s}} and {Merritt}, David},
        title = "{Formation of Galactic Nuclei}",
      journal = {\apj},
         year = 2001,
        month = dec,
       volume = {563},
       number = {1},
        pages = {34-62},
          doi = {10.1086/323830}
}

@ARTICLE{Mingarelli_2012,
       author = {{Mingarelli}, C.~M.~F. and {Grover}, K. and {Sidery}, T. and {Smith}, R.~J.~E. and {Vecchio}, A.},
        title = "{Observing the Dynamics of Supermassive Black Hole Binaries with Pulsar Timing Arrays}",
      journal = {\prl},
         year = 2012,
        month = aug,
       volume = {109},
       number = {8},
          eid = {081104},
          doi = {10.1103/PhysRevLett.109.081104}
}

@ARTICLE{Mingarelli_2018,
       author = {{Mingarelli}, Chiara M.~F. and {Anderson}, Lauren and {Bedell}, Megan and {Spergel}, David N. and {Moran}, Abigail},
        title = "{Improving Binary Millisecond Pulsar Distances with Gaia}",
         year = 2018,
        month = dec,
          eid = {arXiv:1812.06262},
        pages = {arXiv:1812.06262},
          doi = {10.48550/arXiv.1812.06262}
}

@ARTICLE{Petiteau_2013,
       author = {{Petiteau}, Antoine and {Babak}, Stanislav and {Sesana}, Alberto and {de Ara{\'u}jo}, Mariana},
        title = "{Resolving multiple supermassive black hole binaries with pulsar timing arrays. II. Genetic algorithm implementation}",
      journal = {\prd},
         year = 2013,
        month = mar,
       volume = {87},
       number = {6},
          eid = {064036},
        pages = {064036},
          doi = {10.1103/PhysRevD.87.064036},
archivePrefix = {arXiv},
       eprint = {1210.2396},
 primaryClass = {astro-ph.CO},
       adsurl = {https://ui.adsabs.harvard.edu/abs/2013PhRvD..87f4036P}
}

@ARTICLE{Petrov_2025,
       author = {{Petrov}, Polina and {Schult}, Levi and {Taylor}, Stephen R. and {Pol}, Nihan and {Laal}, Nima and {Charisi}, Maria and {Ma}, Chung-Pei},
        title = "{Expectations for the first supermassive black-hole binary resolved by PTAs II: Milestones for binary characterization}",
      journal = {arXiv e-prints},
         year = 2025,
        month = oct,
          eid = {arXiv:2510.01316},
        pages = {arXiv:2510.01316},
          doi = {10.48550/arXiv.2510.01316},
archivePrefix = {arXiv},
       eprint = {2510.01316}
}

@ARTICLE{Phinney_2001,
       author = {{Phinney}, E.~S.},
        title = "{A Practical Theorem on Gravitational Wave Backgrounds}",
      journal = {arXiv e-prints},
         year = 2001,
        month = aug,
          eid = {astro-ph/0108028},
          doi = {10.48550/arXiv.astro-ph/0108028}
}

@ARTICLE{Rajagopal_1995,
    author = {{Rajagopal}, Mohan and {Romani}, Roger W.},
    title = "{Ultra--Low-Frequency Gravitational Radiation from Massive Black Hole Binaries}",
    journal = apj,
    year = 1995,
    month = jun,
    volume = {446},
    pages = {543},
    doi = {10.1086/175813}
}

@ARTICLE{Ravi_2012,
       author = {{Ravi}, V. and {Wyithe}, J.~S.~B. and {Hobbs}, G. and {Shannon}, R.~M. and {Manchester}, R.~N. and {Yardley}, D.~R.~B. and {Keith}, M.~J.},
        title = "{Does a ``Stochastic'' Background of Gravitational Waves Exist in the Pulsar Timing Band?}",
      journal = {\apj},
         year = 2012,
        month = dec,
       volume = {761},
       number = {2},
        pages = {84},
          doi = {10.1088/0004-637X/761/2/84}
}

@ARTICLE{RomanoAllen_2024,
       author = {{Romano}, J.~D. and {Allen}, B.},
        title = "{Answers to frequently asked questions about the pulsar timing array Hellings and Downs curve}",
      journal = {Classical and Quantum Gravity},
         year = 2024,
        month = sep,
       volume = {41},
       number = {17},
          eid = {175008},
        pages = {175008},
          doi = {10.1088/1361-6382/ad4c4c},
archivePrefix = {arXiv},
       eprint = {2308.05847},
 primaryClass = {gr-qc},
       adsurl = {https://ui.adsabs.harvard.edu/abs/2024CQGra..41q5008R}
}

@article{Rosado_2015,
    author = "P A Rosado and Alberto Sesana and Jonathan Gair",
    title = "Expected properties of the first gravitational wave signal detected with pulsar timing arrays",
    journal = mnras,
    volume = 451,
    pages = "2417",
    year = 2015,
    doi = {10.1093/mnras/stv1098}
}

@ARTICLE{Sazhin_1978,
       author = {{Sazhin}, M.~V.},
        title = "{Opportunities for detecting ultralong gravitational waves}",
      journal = {\sovast},
         year = 1978,
        month = feb,
       volume = {22},
        pages = {36-38},
       adsurl = {https://ui.adsabs.harvard.edu/abs/1978SvA....22...36S}
}

@ARTICLE{Sesana_2004,
    author = {{Sesana}, Alberto and {Haardt}, Francesco and {Madau}, Piero and {Volonteri}, Marta},
    title = "{Low-Frequency Gravitational Radiation from Coalescing Massive Black Hole Binaries in Hierarchical Cosmologies}",
    journal = {\apj},
    year = 2004,
    month = aug,
    volume = {611},
    number = {2},
    pages = {623-632},
    doi = {10.1086/422185}
}

@ARTICLE{SesanaVecchioVolonteri_2009,
       author = {{Sesana}, A. and {Vecchio}, A. and {Volonteri}, M.},
        title = "{Gravitational waves from resolvable massive black hole binary systems and observations with Pulsar Timing Arrays}",
      journal = {\mnras},
         year = 2009,
        month = apr,
       volume = {394},
       number = {4},
        pages = {2255-2265},
          doi = {10.1111/j.1365-2966.2009.14499.x},
archivePrefix = {arXiv},
       eprint = {0809.3412},
 primaryClass = {astro-ph},
       adsurl = {https://ui.adsabs.harvard.edu/abs/2009MNRAS.394.2255S}
}

@ARTICLE{SesanaVecchio_2010,
       author = {{Sesana}, Alberto and {Vecchio}, Alberto},
        title = "{Measuring the parameters of massive black hole binary systems with pulsar timing array observations of gravitational waves}",
      journal = {\prd},
         year = 2010,
        month = may,
       volume = {81},
       number = {10},
        pages = {104008},
          doi = {10.1103/PhysRevD.81.104008},
archivePrefix = {arXiv},
       eprint = {1003.0677},
 primaryClass = {astro-ph.CO},
       adsurl = {https://ui.adsabs.harvard.edu/abs/2010PhRvD..81j4008S}
}

@ARTICLE{Simon_2014,
       author = {{Simon}, Joseph and {Polin}, Abigail and {Lommen}, Andrea and {Stappers}, Ben and {Finn}, Lee Samuel and {Jenet}, F.~A. and {Christy}, B.},
        title = "{Gravitational Wave Hotspots: Ranking Potential Locations of Single-source Gravitational Wave Emission}",
      journal = {\apj},
         year = 2014,
        month = mar,
       volume = {784},
       number = {1},
          eid = {60},
        pages = {60},
          doi = {10.1088/0004-637X/784/1/60}
}

@ARTICLE{Speri_2023,
       author = {{Speri}, Lorenzo and {Porayko}, Nataliya K. and {Falxa}, Mikel and {Chen}, Siyuan and {Gair}, Jonathan R. and {Sesana}, Alberto and {Taylor}, Stephen R.},
        title = "{Quality over quantity: Optimizing pulsar timing array analysis for stochastic and continuous gravitational wave signals}",
      journal = {\mnras},
         year = 2023,
        month = jan,
       volume = {518},
       number = {2},
        pages = {1802-1817},
          doi = {10.1093/mnras/stac3237}
}

@ARTICLE{Spiewak_2022,
       author = {{Spiewak}, R. and {Bailes}, M. and {Miles}, M.~T. and {Parthasarathy}, A. and {Reardon}, D.~J. and {Shamohammadi}, M. and {Shannon}, R.~M. and {Bhat}, N.~D.~R. and {Buchner}, S. and {Cameron}, A.~D. and {Camilo}, F. and {Geyer}, M. and {Johnston}, S. and {Karastergiou}, A. and {Keith}, M. and {Kramer}, M. and {Serylak}, M. and {van Straten}, W. and {Theureau}, G. and {Venkatraman Krishnan}, V.},
        title = "{The MeerTime Pulsar Timing Array: A census of emission properties and timing potential}",
      journal = {\pasa},
         year = 2022,
        month = jul,
       volume = {39},
        pages = {e027},
          doi = {10.1017/pasa.2022.19}
}

@ARTICLE{Taylor_2014,
       author = {{Taylor}, Stephen and {Ellis}, Justin and {Gair}, Jonathan},
        title = "{Accelerated Bayesian model-selection and parameter-estimation in continuous gravitational-wave searches with pulsar-timing arrays}",
      journal = {\prd},
         year = 2014,
        month = nov,
       volume = {90},
       number = {10},
          eid = {104028},
          doi = {10.1103/PhysRevD.90.104028},
       eprint = {1406.5224}
}

@ARTICLE{Titarchuk_2023,
       author = {{Titarchuk}, Lev and {Seifina}, Elena and {Shrader}, Chris},
        title = "{OJ 287: A new BH mass estimate of the secondary}",
      journal = {\aap},
         year = 2023,
        month = mar,
       volume = {671},
          eid = {A159},
        pages = {A159},
          doi = {10.1051/0004-6361/202345923},
archivePrefix = {arXiv},
       eprint = {2302.06068}
}

@ARTICLE{vanHaasteren2009,
       author = {{van Haasteren}, Rutger and {Levin}, Yuri and {McDonald}, Patrick and {Lu}, Tingting},
        title = "{On measuring the gravitational-wave background using Pulsar Timing Arrays}",
      journal = {\mnras},
         year = 2009,
        month = may,
       volume = {395},
       number = {2},
        pages = {1005-1014},
          doi = {10.1111/j.1365-2966.2009.14590.x},
archivePrefix = {arXiv},
       eprint = {0809.0791},
 primaryClass = {astro-ph},
       adsurl = {https://ui.adsabs.harvard.edu/abs/2009MNRAS.395.1005V}
}

@ARTICLE{vanHaasterenLevin_2013,
       author = {{van Haasteren}, Rutger and {Levin}, Yuri},
        title = "{Understanding and analysing time-correlated stochastic signals in pulsar timing}",
      journal = {\mnras},
         year = 2013,
        month = jan,
       volume = {428},
       number = {2},
        pages = {1147-1159},
          doi = {10.1093/mnras/sts097},
archivePrefix = {arXiv},
       eprint = {1202.5932},
 primaryClass = {astro-ph.IM},
       adsurl = {https://ui.adsabs.harvard.edu/abs/2013MNRAS.428.1147V}
}

@ARTICLE{Vigeland_2018,
    author = {{Vigeland}, Sarah J. and {Islo}, Kristina and {Taylor}, Stephen R. and {Ellis}, Justin A.},
    title = "{Noise-marginalized optimal statistic: A robust hybrid frequentist-Bayesian statistic for the stochastic gravitational-wave background in pulsar timing arrays}",
    journal = prd,
    year = 2018,
    month = aug,
    volume = {98},
    pages = {044003},
    doi = {10.1103/PhysRevD.98.044003}
}

@ARTICLE{Wang_2017,
       author = {{Wang}, Yan and {Mohanty}, Soumya D.},
        title = "{Pulsar Timing Array Based Search for Supermassive Black Hole Binaries in the Square Kilometer Array Era}",
      journal = {\prl},
         year = 2017,
        month = apr,
       volume = {118},
       number = {15},
        pages = {151104},
          doi = {10.1103/PhysRevLett.118.151104}
}

@ARTICLE{Zhu_2016,
       author = {{Zhu}, X. -J. and {Wen}, L. and {Xiong}, J. and {Xu}, Y. and {Wang}, Y. and {Mohanty}, S.~D. and {Hobbs}, G. and {Manchester}, R.~N.},
        title = "{Detection and localization of continuous gravitational waves with pulsar timing arrays: the role of pulsar terms}",
      journal = {\mnras},
         year = 2016,
        month = sep,
       volume = {461},
       number = {2},
        pages = {1317-1327},
          doi = {10.1093/mnras/stw1446},
archivePrefix = {arXiv},
       eprint = {1606.04539}
}

@misc{enterprise,
    author       = {Justin A. Ellis and Michele Vallisneri and Stephen R. Taylor and Paul T. Baker},
    title        = {ENTERPRISE: Enhanced Numerical Toolbox Enabling a Robust PulsaR Inference SuitE},
    month        = sep,
    year         = 2020,
    howpublished = {Zenodo},
    doi          = {10.5281/zenodo.4059815},
    url          = {https://doi.org/10.5281/zenodo.4059815}
}

@misc{enterprise_extensions,
    author       = {Stephen R. Taylor and Paul T. Baker and Jeffrey S. Hazboun and Joseph Simon and Sarah J. Vigeland},
    title        = {enterprise\_extensions},
    year         = {2021},
    url          = {https://github.com/nanograv/enterprise_extensions},
    note         = {v2.4.3}
}

@software{libstempo,
       author = {{Vallisneri}, Michele},
        title = "{libstempo: Python wrapper for Tempo2}",
 howpublished = {Astrophysics Source Code Library, record ascl:2002.017},
         year = 2020,
        month = feb,
          eid = {ascl:2002.017},
       adsurl = {https://ui.adsabs.harvard.edu/abs/2020ascl.soft02017V}
}

@Article{numpy,
    title         = {Array programming with {NumPy}},
    author        = {Charles R. Harris and K. Jarrod Millman and St{\'{e}}fan J. van der Walt and Ralf Gommers and Pauli Virtanen and David Cournapeau and Eric Wieser and Julian Taylor and Sebastian Berg and Nathaniel J. Smith and Robert Kern and Matti Picus and Stephan Hoyer and Marten H. van Kerkwijk and Matthew Brett and Allan Haldane and Jaime Fern{\'{a}}ndez del R{\'{i}}o and Mark Wiebe and Pearu Peterson and Pierre G{\'{e}}rard-Marchant and Kevin Sheppard and Tyler Reddy and Warren Weckesser and Hameer Abbasi and Christoph Gohlke and Travis E. Oliphant},
    year          = {2020},
    month         = sep,
    journal       = {Nature},
    volume        = {585},
    number        = {7825},
    pages         = {357--362},
    doi           = {10.1038/s41586-020-2649-2},
    publisher     = {Springer Science and Business Media {LLC}},
    url           = {https://doi.org/10.1038/s41586-020-2649-2}
}

@Article{matplotlib,
    Author    = {Hunter, J. D.},
    Title     = {Matplotlib: A 2D graphics environment},
    Journal   = {Computing in Science \& Engineering},
    Volume    = {9},
    Number    = {3},
    Pages     = {90--95},
    publisher = {IEEE COMPUTER SOC},
    doi       = {10.1109/MCSE.2007.55},
    year      = 2007
}

@article{astropy:2013,
    author = {{Astropy Collaboration} and {Robitaille}, T.~P. and {Tollerud}, E.~J. and {Greenfield}, P. and {Droettboom}, M. and {Bray}, E. and {Aldcroft}, T. and {Davis}, M. and {Ginsburg}, A. and {Price-Whelan}, A.~M. and {Kerzendorf}, W.~E. and {Conley}, A. and {Crighton}, N. and {Barbary}, K. and {Muna}, D. and {Ferguson}, H. and {Grollier}, F. and {Parikh}, M.~M.},
    doi = {10.1051/0004-6361/201322068},
    eprint = {1307.6212},
    journal = {\aap},
    month = oct,
    pages = {A33},
    primaryclass = {astro-ph.IM},
    title = {{Astropy: A community Python package for astronomy}},
    volume = 558,
    year = 2013
}

@ARTICLE{astropy:2018,
    author = {{Astropy Collaboration} and {Price-Whelan}, A.~M. and
         {Sip{\H{o}}cz}, B.~M. and {G{\"u}nther}, H.~M. and {Lim}, P.~L. and
         {Crawford}, S.~M. and {Conseil}, S. and {Shupe}, D.~L. and
         {Craig}, M.~W. and {Dencheva}, N. and {Ginsburg}, A. and {Vand
        erPlas}, J.~T. and {Bradley}, L.~D.},
    title = "{The Astropy Project: Building an Open-science Project and Status of the v2.0 Core Package}",
    journal = {\aj},
    year = 2018,
    month = sep,
    volume = {156},
    number = {3},
    pages = {123},
    doi = {10.3847/1538-3881/aabc4f}
}

@ARTICLE{astropy:2022,
    author = {{Astropy Collaboration} and {Price-Whelan}, Adrian M. and {Lim}, Pey Lian and {Earl}, Nicholas and {Starkman}, Nathaniel and {Bradley}, Larry and {Shupe}, David L. and {Patil}, Aarya A. and {Corrales}, Lia and {Brasseur}, C.~E. and {N{"o}the}, Maximilian and {Donath}, Axel and {Tollerud}, Erik and {Morris}, Brett M.},
    title = "{The Astropy Project: Sustaining and Growing a Community-oriented Open-source Project and the Latest Major Release (v5.0) of the Core Package}",
    journal = {\apj},
    year = 2022,
    month = aug,
    volume = {935},
    number = {2},
    pages = {167},
    doi = {10.3847/1538-4357/ac7c74}
}

@ARTICLE{tempo2,
	author = {{Hobbs}, G.~B. and {Edwards}, R.~T. and {Manchester}, R.~N.},
	title = "{TEMPO2, a new pulsar-timing package - I. An overview}",
	journal = {\mnras},
	year = 2006,
	month = jun,
	volume = {369},
	number = {2},
	pages = {655-672},
	doi = {10.1111/j.1365-2966.2006.10302.x},
	archivePrefix = {arXiv},
	eprint = {astro-ph/0603381},
	primaryClass = {astro-ph},
	adsurl = {https://ui.adsabs.harvard.edu/abs/2006MNRAS.369..655H}
}

@article{Lentati2014,
	doi = {10.1093/mnras/stt2122},
	url = {https://doi.org/10.1093\%2Fmnras\%2Fstt2122},
	year = 2014,
	month = {dec},
	publisher = {Oxford University Press ({OUP})},
	volume = {437},
	number = {3},
	pages = {3004--3023},
	author = {L. Lentati and P. Alexander and M. P. Hobson and F. Feroz and R. van Haasteren and K. J. Lee and R. M. Shannon},
	title = {temponest: a Bayesian approach to pulsar timing analysis},
	journal = {\mnras}
}

@ARTICLE{chainconsumer,
    author = {{Hinton}, S.~R.},
    title = "{ChainConsumer}",
    journal = {The Journal of Open Source Software},
    year = 2016,
    month = aug,
    volume = 1,
    pages = {00045},
    doi = {10.21105/joss.00045},
    adsurl = {http://adsabs.harvard.edu/abs/2016JOSS....1...45H}
}

@misc{ptmcmc,
  author       = {Justin Ellis and
                  Rutger van Haasteren},
  title        = {jellis18/PTMCMCSampler: Official Release},
  month        = oct,
  year         = 2017,
  doi          = {10.5281/zenodo.1037579},
  url          = {https://doi.org/10.5281/zenodo.1037579}
}

\end{document}